\documentclass[12pt,preprint]{aastex}

\slugcomment{ApJ, accepted  07 May 2003}
\shorttitle{Gravity indicators in the BD spectra}
\shortauthors{Gorlova et al.}

\begin{document}

\title{Gravity Indicators in the Near-Infrared Spectra of Brown Dwarfs}

\author{N. I. Gorlova, M. R. Meyer, G. H. Rieke, J. Liebert}
\affil{Steward Observatory, The University of Arizona, 933 N. Cherry Ave., 
Tucson, AZ  85721}
\email{ngorlova@as.arizona.edu, mmeyer@as.arizona.edu}

\begin{abstract}
We investigate the sensitivity to temperature and gravity of the strong 
absorption features in the J- and K-band spectra of substellar objects. We compare the
spectra of giants and young M dwarfs (of low gravity) to field M and L dwarfs (of high gravity)
and to model spectra from the Lyon group. We find that low-resolution spectra of M4 -- M9 stars and
young brown dwarfs at R $\sim$ 350 and $S/N > 70$ can determine the spectral type to a 
precision of $\pm$ 1 subtype, using the H$_2$O and CO bands, and can measure the surface gravity
to $\pm$ 0.5 dex, using the atomic lines of \ion{K}{1} and \ion{Na}{1}. This result points toward the
development of photometric spectral indices to separate low-mass members from 
foreground and background objects
in young clusters and associations. We also emphasize the complexity of the interpretation
of the empirical quantities (e.g., spectral types) in terms of the physical variables (e.g.,
temperature, opacities) in the cool atmospheres of young brown dwarfs.

\end{abstract}


\section{Introduction}

One of the central problems in stellar astrophysics is the shape of the initial mass function (IMF)
at the sub-stellar end. Is it terminated somewhere below the hydrogen burning limit? 
Is it influenced by the conditions of the ambient medium? These questions (and others) are
prompted by the theoretical speculations on opacity-limited fragmentation  
\citep{Bate02,Boss01} and the boundary between
the structural properties of stars, brown dwarfs (BDs), and planets \citep[e.g.,][]{BurrowsRMP01}. 
In addition to studies aimed at determining the time- and space-averaged
field star IMF \citep[e.g.,][]{Reid99,Kroupa01}, it is interesting to compare
results from young stars to the field to search for variations in the IMF. 
Evidence to date suggests that the stellar IMF ($>$ 0.08 M$_{\odot}$) in young
clusters is  consistent with results from the field \citep{Meyer00}. 
Previous results suggest that a universal IMF may extend into the sub-stellar 
regime \citep{Luhman00,Martin00}. 
However, \citet{Briceno02} find evidence for a difference in the sub-stellar IMF
between the T association in the Taurus-Auriga dark cloud and the dense
Trapezium cluster associated with the Orion Nebula.  It is still unclear 
whether the IMF truncates near the minimum mass defined by opacity-limited
fragmentation in the interstellar medium \citep{Spitzer78}. 

Observational approaches to answering these questions 
are hampered by the fading of the low-mass objects as they
age, while the nearest birth places (where they are young and
bright) are a few hundred parsecs away and often highly extincted. Thus, the use of large
telescopes and infrared instrumentation is required, and even then sensitivity limitations
can make determination of fundamental parameters or even cluster membership difficult. 
For example, a common way to determine the IMF in young clusters is by means of broad-band photometry. 
However, though fast, this method suffers from contamination by foreground dwarfs and background
giants, which can have the same colors and apparent magnitudes as the pre-main-sequence (PMS)
population at the very lowest masses. 

Another way to identify young objects is to search for signs of extreme activity. The
presence of accretion disks is indicated by photometric excesses \citep{Natta01, Muench01, Natta02, Liu03}, sometimes accompanied by line emission and veiling \citep{Luhman98a, Gomez02, ZapOs02a, ZapOs02b}. Young objects may also exhibit photometric
variability due to modulation by rotating spots \citep{Joergens02}. 
Magnetic activity may be manifested in strong X-ray emission and flares \citep{Imanishi01, Feigelson02, Preibisch02}. Unfortunately, the biases inherent in
samples defined by these signatures of activity are not well understood. 

Another approach is to conduct spectroscopic follow-up of photometric candidates. Unfortunately,
conventional R $\sim$ 1000 -- 2000 spectroscopy is very time-consuming in the infrared. An intermediate
approach would be to find a set of strong features, detectable  at low resolution or even by
means of narrow-band photometry, that are capable of discriminating between various types
of red objects in the visible \citep{Hillenbrand02,Clarke02}. 
To extend this method further into the infrared, we have searched for gravity-sensitive 
features in J- and K-band spectra. Young brown dwarfs 
(and low-mass stars) should have surface gravities intermediate between dwarfs and giants. 
We have obtained near-infrared spectra of confirmed members of young open clusters 
(low gravity) and of field objects (high gravity). We combine photometric and visible 
spectroscopic data from the literature 
with theoretical evolutionary tracks in order to estimate the luminosities, masses, and 
gravities of these objects. We then attempt to disentangle the effects of 
temperature and gravity on the infrared spectral features of these objects, 
using model spectra for guidance. 

\section{New Observations}
\subsection{Sample Selection}

Since our goal is to identify gravity-sensitive spectral features, we selected a sample
of objects with a narrow range of spectral types (SpTs), thus removing temperature so
far as possible as a variable in the data. Low-mass objects initially contract at nearly
constant temperature; thus, by selecting sources from young clusters with ages from
$\le$ 1 to $>$ 100 Myr we can probe gravities differing by nearly two orders of magnitude
in objects of similar mass! Although field M dwarfs are much older, their gravities are only
slightly larger than those of the oldest open cluster objects in our sample. However, they have
the advantage of being much closer, and therefore much brighter than the BDs in old clusters.

We therefore included three types of objects in our sample, of SpT M4 -- M9 and of mass $\sim 0.1 M_\odot$: 
1.) Members of the star-forming regions Taurus, $\rho$ Ophiuchi, IC 348, $\sigma$ Orionis, 
Upper Scorpius, and TW Hydrae; 2.) one object from the Pleiades (age $\sim$ 120 Myr \citep{Stauffer98}); and 3.) field dwarfs (presumed to be more than 500 Myr old) selected
to match the SpTs of our cluster objects, with a few additional later-type sources. We
also observed three late M giants to calibrate the low gravity end of the scale. The program
objects are listed in Table~\ref{litr}.

\subsection{Data Acquisition}\label{datred}

We observed the sample with the near-infrared spectrometer FSpec \citep{Williams93} on the 6.5m
Monolithic Mirror Telescope. Table~\ref{obs} is a log of the observations. Source acquisition and guiding
were performed with a slit-viewing infrared camera. The width of the slit was 1.0 arcsec 
($\sim$ 3 pixels) and we used a 75 l/mm grating. After Gaussian smoothing with $\sigma =$ 1 pixel,
the final resolution of the spectra was 0.0045$\mu$m (or R $\sim$ 300) in J-band and 
0.0065$\mu$m (or R $\sim$ 350) in K-band. The useful spectral coverage was 1.16 -- 1.34 and 
2.10 -- 2.41$\mu$m, respectively. Each observation consisted of four frames, each at
a different position along the 30$\arcsec$ slit. The typical exposure time was 120s per frame.
The giants were observed on a cloudy night to avoid saturation. Each target observation was preceded and
followed by an identical spectrum (except for integration time) of a field G or F dwarf, 
selected to be within 0.3 airmasses, to allow correction for telluric absorption.  
  
\begin{figure}
\epsscale{1.0}
\plotone{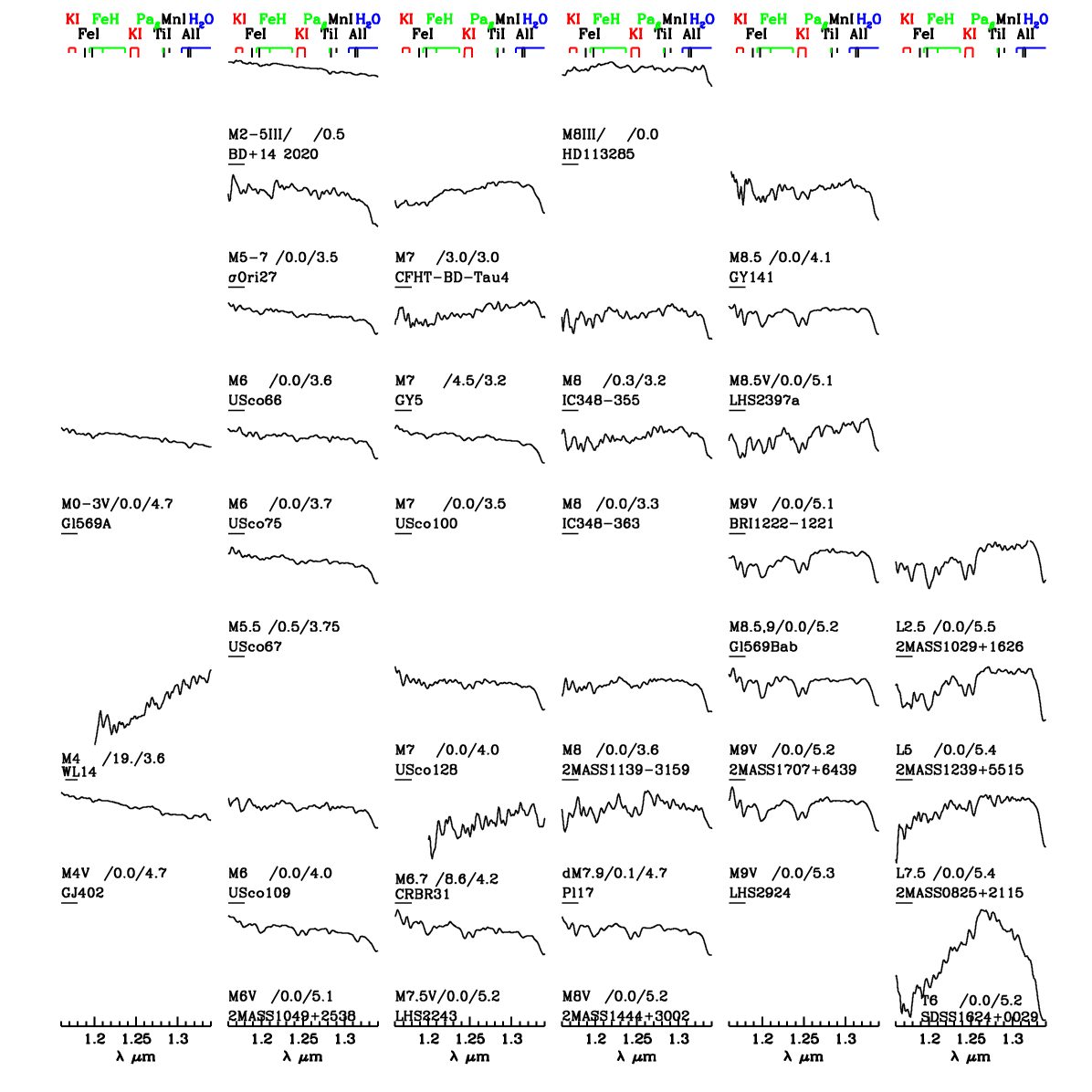}
\caption{\label{jsp}
FSpec J-band spectra of program objects, arranged according to SpT in the horizontal
direction and log(g) in the vertical. Zero-flux level is shown as a small dash 
below an object name. The abbreviation */*/* stands for SpT / A$_{V}$ / log(g).
Some of the lines identified on top may be absent in giants.
}
\end{figure}

\begin{figure}
\epsscale{1.0}
\plotone{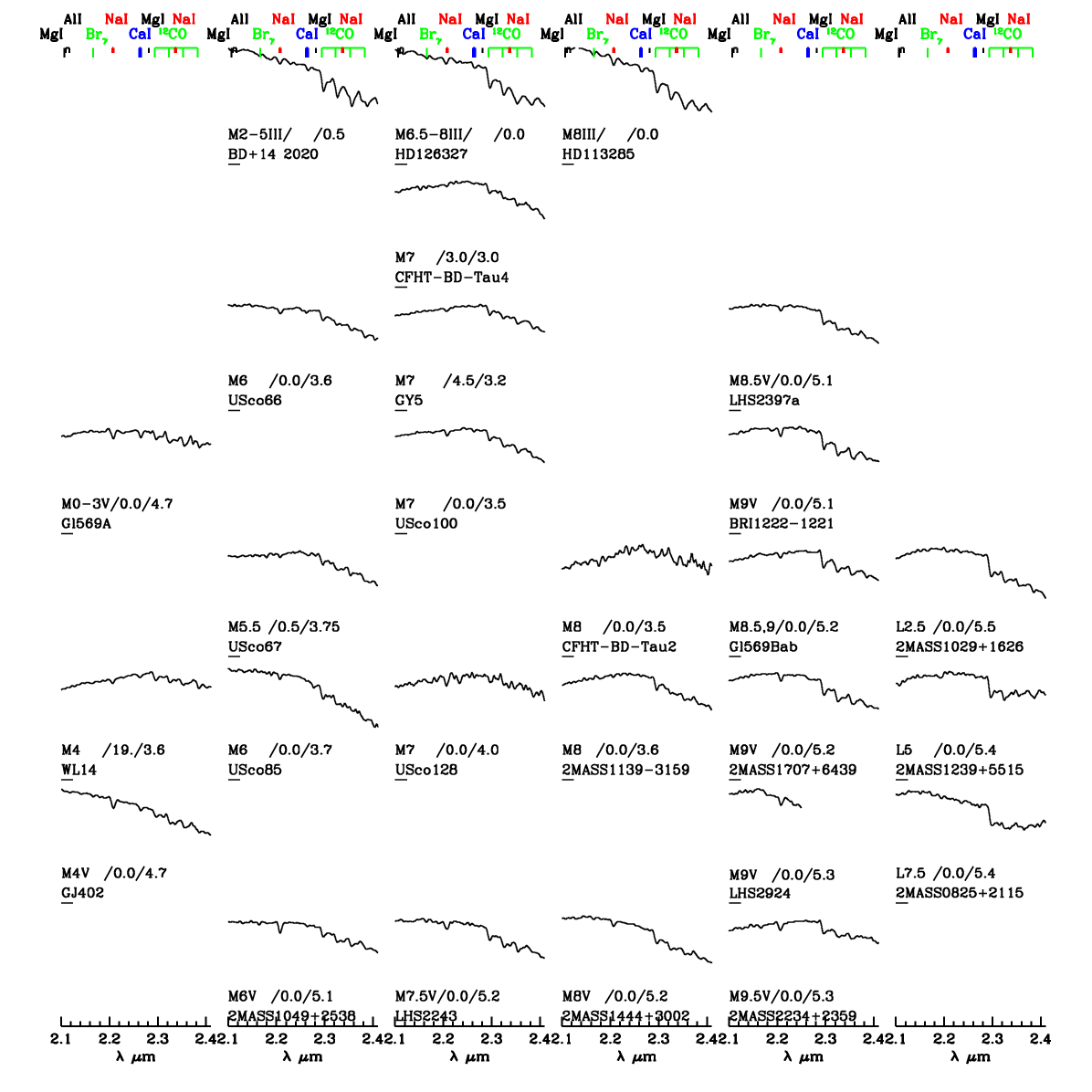}
\caption{\label{ksp}
FSpec K band spectra. Same designations as in Fig.~\ref{jsp}. 
}
\end{figure}

Spectral extraction used standard IRAF tasks as well as specific routines for FSpec provided
by C. Engelbracht. First, the images were dark-subtracted and bad pixels were marked (using a
mask created for each night by combining flat and dark frames). Then the frames were
differenced, assigning unequal weights as necessary for better airglow removal. They were 
divided by a dome flat and rectified to correct for the geometric distortion 
of the spectrograph. Finally,
they were median-combined (which also removed nearly all the bad pixels). We also used the
data to obtain a pure airglow spectrum, which was used for wavelength calibration.
After extracting an object spectrum, it was divided by a similarly reduced telluric standard spectrum 
and the result was multiplied by a normalized solar spectrum to correct for absorption lines in
the standard. We multiplied the resultant spectrum by a blackbody spectrum of temperature 
corresponding to the SpT of the standard and finally smoothed the result in the IRAF task gauss
with the sigma parameter set to one pixel. No absolute calibration was attempted; all the
spectra were normalized to the same scale based on averages over the spectral regions
1.185 - 1.305$\mu$m or 2.100 - 2.340$\mu$m. The spectra are shown in Figures~\ref{jsp}, \ref{ksp}. 

\section{Source Properties}
\subsection{Properties of PMS Sub-giants and Field Dwarf Stars}

\subsubsection{Surface Gravities}

Although it is clear qualitatively that younger objects have lower gravity, quantitative
analysis requires estimates of log(g) that rely in part on theoretical evolutionary models.
Thus, to further analyze the spectral features we need to determine the surface gravities of the objects. 

We estimated the gravities according to

\begin{equation} 
log(g) = 4.42 + log\left(\frac{M}{M_{\odot}} \right) - log\left(\frac{L}{L_{\odot}} \right) + 4\times log\left(\frac{T_{eff}}{5770} \right) 
\label{loggeqn}
\end{equation}

\noindent
Temperatures were obtained from the spectral types, luminosities were based on bolometric 
corrections to dereddened I- or J-band magnitudes, and masses were determined from 
theoretical tracks on the L -- T$_{eff}$ diagram as described below. 
Relevant references to the literature
are included in Table~\ref{litr}, and the source properties we derive are listed in Table~\ref{thisw}. The locations 
of the sample objects in the H-R diagram for a subset of tracks and one temperature scale are 
illustrated in Figure~\ref{hrD}. 

\subsubsection{Luminosities}

Luminosities were calculated according to:

\begin{equation}
log\left(\frac{L}{L_{\odot}} \right) = 1.86 - 0.4\times \left(J-0.265\times A_{V} + BC_{J} + 5 - 5\times log(d)\right)
\label{logleqn}
\end{equation}

\noindent
This equation assumes M$_{bol\odot}$ = 4.64 (BC$_{V\odot}$ = $-$ 0.19, \citet[][p.60]{Binney98}) and A$_J$ = 0.265~A$_V$ (\citet{Cohen81}), J in the CIT system.

The J band is optimal for luminosity estimation in young low-mass objects for several reasons 
\citep{Meyer97, Luhman99, Natta02}: 1.) most of our sources
are not heavily reddened and at temperatures of 2000 -- 3000K, so the peak of their spectral energy
distributions falls within this band; and 2.) the J band is also situated in the range least
contaminated by UV and IR excesses due to possible accretion disks or other forms of circumstellar material.
For a few USco objects missing J magnitudes, we
estimated them from I magnitudes by combining the (I -- J) colors of USco objects of
similar SpTs from \citet{Ardila00} with the field dwarf SpT -- (I -- J)
relationship from \citet[][their Fig.4]{Dahn02}, adopting (I -- J) to be 2.15, 2.20, and 2.40
magnitudes for M5.5, M6, and M7 respectively. For the hottest star in our sample (Gl 569A), the
bolometric magnitude was derived from m$_V$ and BC$_V$ (assumed to be -- 1.49 for M1.5), since no
infrared photometry is available for this star.  

The extinction for cluster objects was taken from the literature, 
where it was usually derived by de-reddening the sources using intrinsic colors inferred from 
known spectral types. 
Since the true colors of young BDs are not yet well-established, field dwarf colors
were used for reddening determination. We later justify this procedure by showing that 
these young objects are much closer in spectral properties to dwarfs than to giants and that the features
used in the optical spectral classification (TiO, VO) are not very sensitive to gravity for
temperatures higher than $\sim$ 2500K. For field dwarfs thought to be within 25 parsecs of the Sun, 
we assumed A$_V$ = 0.0.

Bolometric corrections (BC$_J$) were calculated by averaging SpT vs. BC$_J$ data from three sources: 
\citet{WGM99} for M dwarfs, \citet{Reid01} for L dwarfs, and \citet{Dahn02}
for both M and L dwarfs.  

Our distance estimates are based on the HIPPARCOS convergent-point and parallax results
\citep{Perryman97, deZeeuw99} and those by \citet{Monet92} and \citet{Dahn02}. The only exception is the
Pleiades, where we adopted recent result from main sequence fitting \citep{StNissen01},
and incorporated the difference between two independent estimates (0.25 mag) into the 
distance modulus error. For 2MASSJ1139-3159, a member of the TW Hydra association,
the distance is comparable with the size of association itself, and is therefore very
uncertain. The distance uncertainty was included in estimating the effects of observational errors on the
physical properties in Section~\ref{err}. For the seven field dwarfs with unknown distances,
we adopted M$_J$ from the M$_J$ vs. SpT relation of \citet{Dahn02}. For the only
T dwarf in our sample (SDSS 1624+0029, T6), we adopted luminosity and temperature equal to
those of Gl 229B (T6.5), as given in \citet[][Table 7]{Leggett02}, since no reliable BCs
are available for T dwarfs yet.

\subsubsection{Temperatures}

Effective temperatures of our sources were estimated based on their SpTs. There are two difficulties
in this procedure. First, the dwarf temperature scale is uncertain by $\pm$ 200K as shown
in Figure~\ref{tempfig}, where we compile some recent temperature estimates from the literature. Second,
the same spectral type may not correspond to the same temperature in a dwarf, a PMS object, and a 
giant. Due to the paucity of direct T$_{eff}$ measurements (low-mass eclipsing binaries for linear
diameter measurements are yet to be discovered), most temperature scales are 
based on comparison with models. However, the models may still be inadequate. For example,
\citet{Lucas01} derived SpTs of substellar sources in the Trapezium by comparing
de-reddened H-band spectra to those of field dwarfs. They then used Dusty99 model
spectra by \citet{Allard00} to derive temperatures, which turned out to be very hot 
for their late types compared to the field, even allowing for subdwarf gravities (Figure~\ref{tempfig}).
On the other hand, \citet{Natta02} using the same but slightly updated model spectra
on the low-resolution {\it full} JHK spectra of young BDs in $\rho$ Oph, arrived at a 
dwarf temperature scale. We will comment further on this discrepancy in Section~\ref{optsp}.

Taking into account these uncertainties, we based our temperature estimates on 
observationally based calibrations: two of them \citep[][further WGM99]{Luhman99,WGM99} 
were used to study the low mass population of young clusters, and the other
\citep{Dahn02} is based on the trigonometric parallaxes to field dwarfs. The average
value of the temperature from these three calibrations for each program object is given
in Table~\ref{thisw}.

\subsubsection{Evolutionary Tracks and Masses}\label{evtrm}

There are three state-of-the-art sets of theoretical tracks available for low-mass objects,
differing in the treatment of interior structures and especially atmospheres: 1.) those of
\citet{Burrows97} are well-suited for the lowest mass BDs and planetary-mass objects; 2.)
\citet{dm97, dm98} (DM97, DM98) use gray atmospheres and calculate tracks for very young
objects; and 3.) the Lyon group 
includes constraints from detailed atmospheric codes --
dust-free ``NextGen'' \citep{nga,ngb} and more realistic at cooler temperatures ``Dusty'' \citep{Allard01} --
to construct BCAH98 \citep{bcah98} and Dusty00 \citep{chabrier00} evolutionary tracks, respectively.
However, we should bear in mind that none of
the current models is capable of describing adequately ages less than 1 Myr \citep{Baraffe02}.
While many of our cluster objects probably fall within this age range, the resulting
systematic uncertainties in estimating their gravity (which would affect our results) 
seem to be $<$ 0.5 dex as described below. 

As one can see from Table~\ref{thisw}, our program objects (except for WL 14, Gl 569A, and GJ 402) lie in
the vicinity of, or below the BD boundary at 0.075 M$_\odot$. For our study it is
not necessary that all the objects be BDs, because the transition from the stellar to the substellar
regime is continuous in terms of luminosity, temperature, and other properties. Table 3 shows that
the average masses of young and old objects in our sample are very close, allowing us to consider
the spectra as an evolutionary sequence.

We adopted the average of the gravities estimated independently from three sets of tracks and three
temperature scales (Figure~\ref{hrall}). As can be seen from Table~\ref{thisw} and Figure~\ref{hrgT}, 
the gravities of
the young objects are well-separated from those of field dwarfs: log(g) = 3.0 -- 4.2 for the former, and
5.0 -- 5.5 for the latter. M giants have much lower gravities than dwarfs of either age
range, log(g) = 0.0 $\pm$ 0.5 (Section~\ref{gts}). 

\subsubsection{Uncertainties in the Physical Parameters}\label{err}
  
There are two major types of uncertainty in the physical parameters in Table~\ref{thisw}: 
1) those systematic uncertainties due to the choice of theoretical tracks and 
the SpT/T$_{eff}$ relations; and 2) random uncertainties due to measurement error 
and uncertainties in calibrations. We consider each in turn.

As explained above, we averaged estimates from several sets of theoretical tracks
rather than giving preference to one of them. The dispersion among these estimates provides
a sense of the uncertainty due to {\it theory} in our results. However, there may also be
systematic errors that affect the results of all the models in similar ways, an issue
that is difficult to treat quantitatively. 

Thus, for each object we have assigned from one to three values of T$_{eff}$ using SpT/T$_{eff}$
relations from \citet{Dahn02, Luhman99, WGM99}. The
average values are given in Table~\ref{thisw} and used in Figure~\ref{hrgT}. The agreement among these temperature
scales is of about 80K for field objects, and 130K for cluster objects. 
There is a parallel observationally
based error due to the uncertainty in SpTs of $\pm$ 1 subtype for cluster objects and half
of this amount for field objects. Including the effects of the dispersion in the \citet{Dahn02} 
SpT/T$_{eff}$ relation of $\pm$ 100K (their Figure 7), we arrive at an uncertainty in T$_{eff}$
due to measurement error of 120K for field dwarfs and 150K for cluster members. These latter
errors are used in the following discussion as they are larger than those from the indicated
range of possible temperature scales.  We note that \citet{Luhman99} finds the typical 
offset between dwarf and subgiant scales is +150K from M6 -- M8, of order the random 
uncertainties adopted here. 

\begin{figure}
\epsscale{1.0}
\plotone{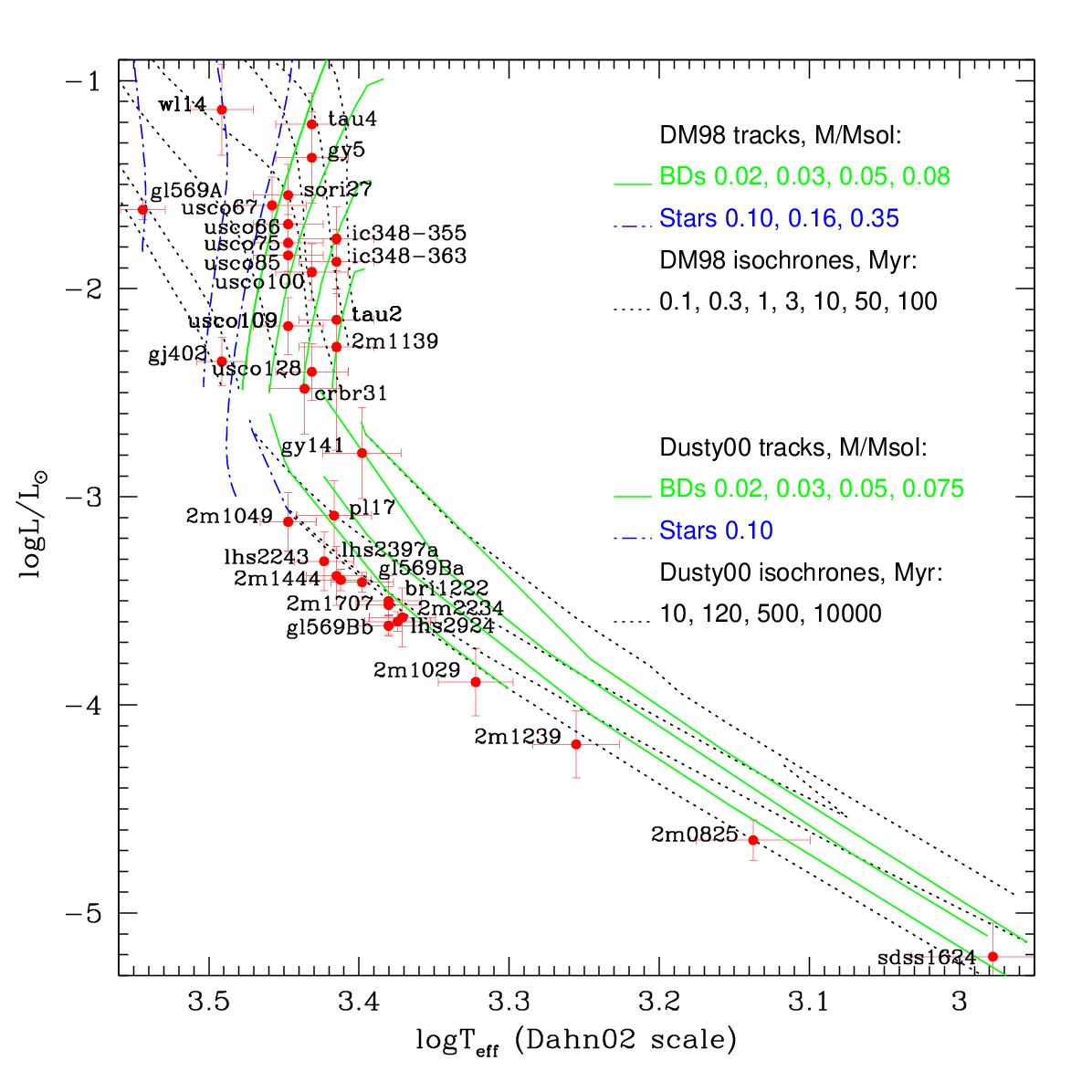}
\caption{\label{hrD}
Program objects (except for giants), overlaid on \citet{dm97,dm98} tracks
(younger and more luminous objects) and on \citet{chabrier00} Dusty tracks. Temperature scale
by \citet{Dahn02} was used in this plot. Masses and ages are defined from right to left (towards increasing 
temperature). The figure is for guidance only to show relative positions 
of our targets on the H-R diagram (see \S~\ref{evtrm}.)
}
\end{figure}

\begin{figure}
\epsscale{1.0}
\plotone{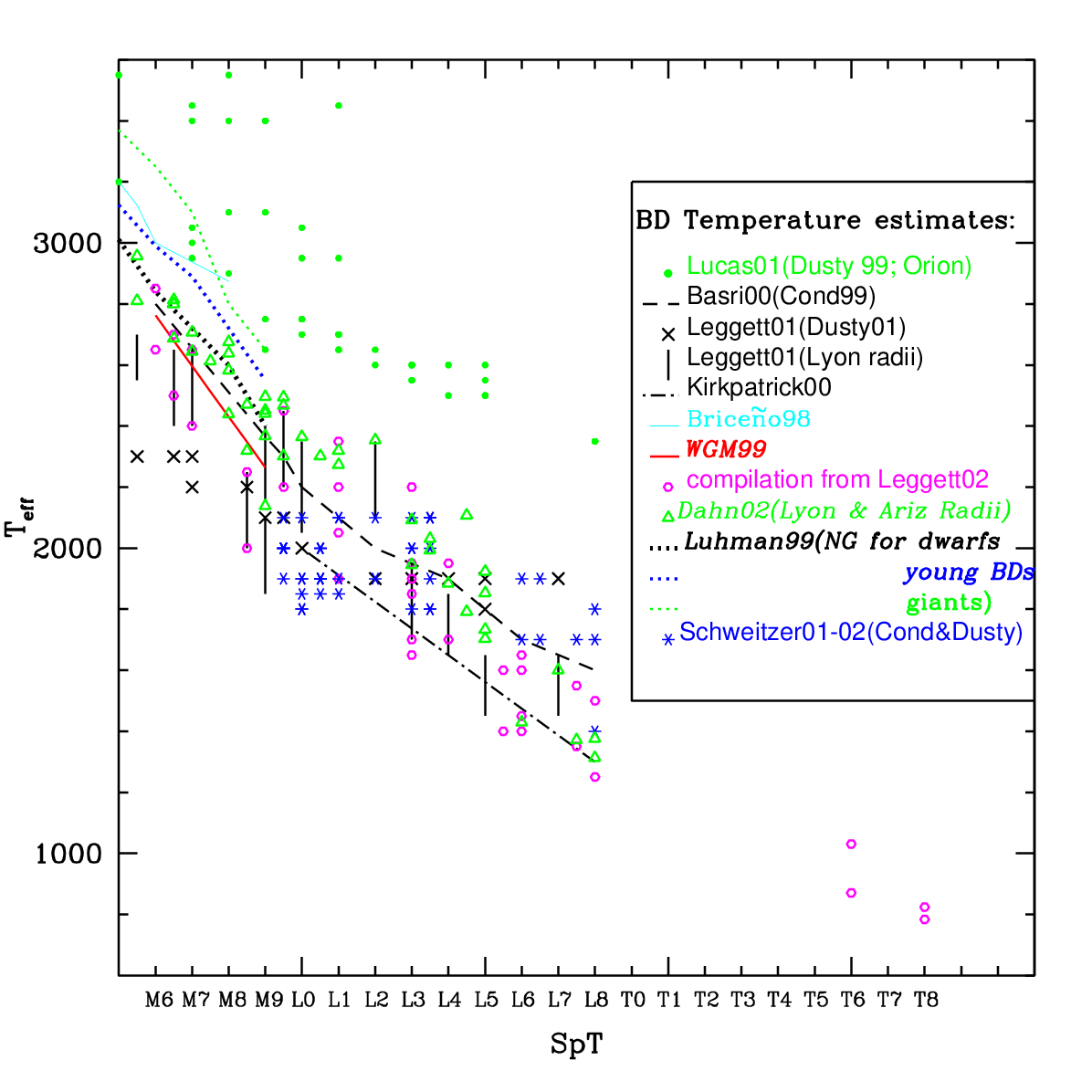}
\caption{\label{tempfig}
Some recent temperature scales drawn from the literature. Most of them were
derived by comparison to model spectra by Lyon group (Cond and Dusty).
The latest one, by \citet{Dahn02} uses parallaxes and theoretical radii.
In this work we considered three calibrations: by WGM99, Dahn02 and Luhman99.
}
\end{figure}

\begin{figure}
\epsscale{1.0}
\plotone{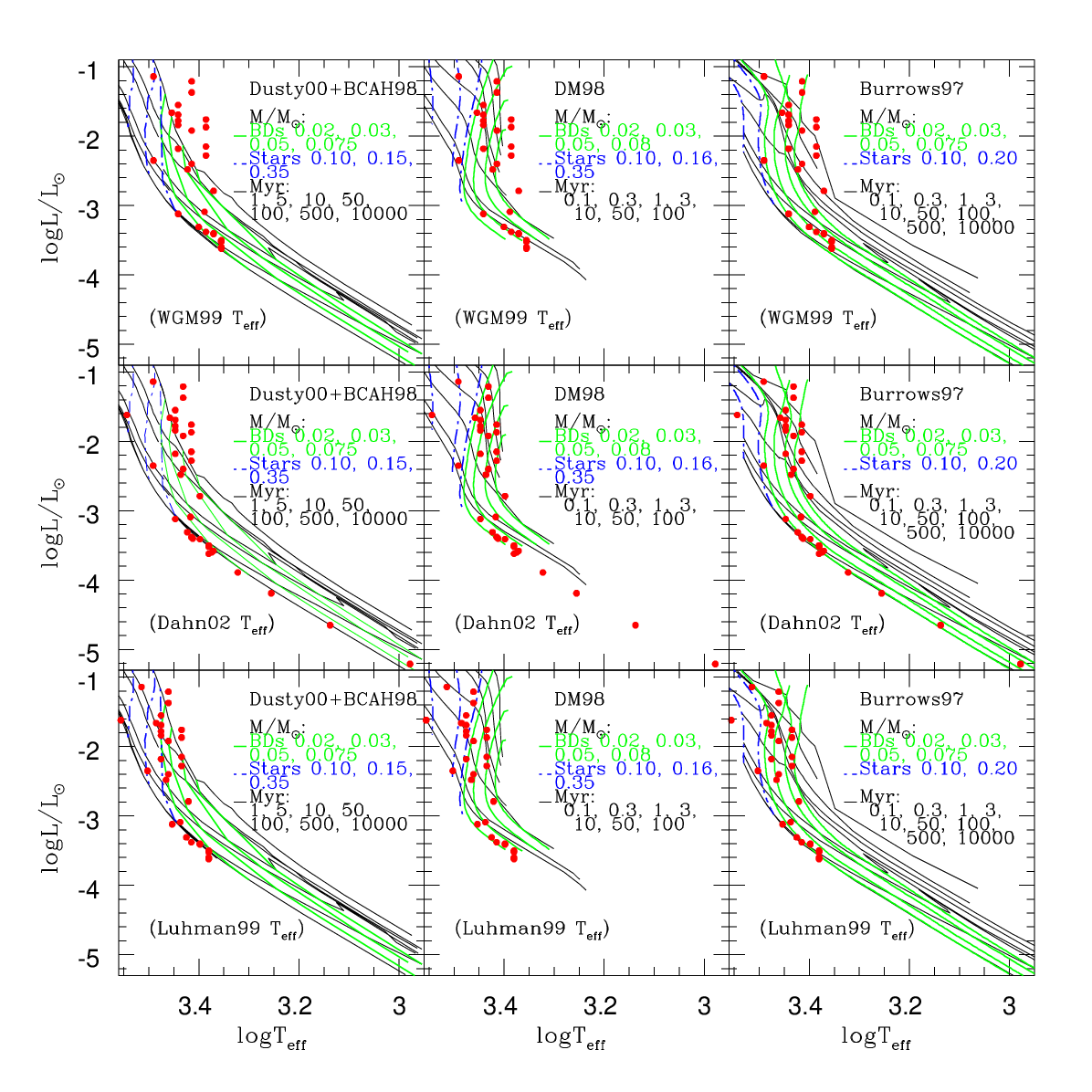}
\caption{\label{hrall}Program objects (dots) superimposed on the 3 sets of evolutionary tracks
using 3 SpT--T$_{eff}$ relations. Similar notation is used for tracks and isochrones
as in Fig.~\ref{hrD}. The figure shows how the choice of tracks and temperature scales
influences derived masses and ages for low-mass objects.
}
\end{figure}

\begin{figure}
\epsscale{1.0}
\plotone{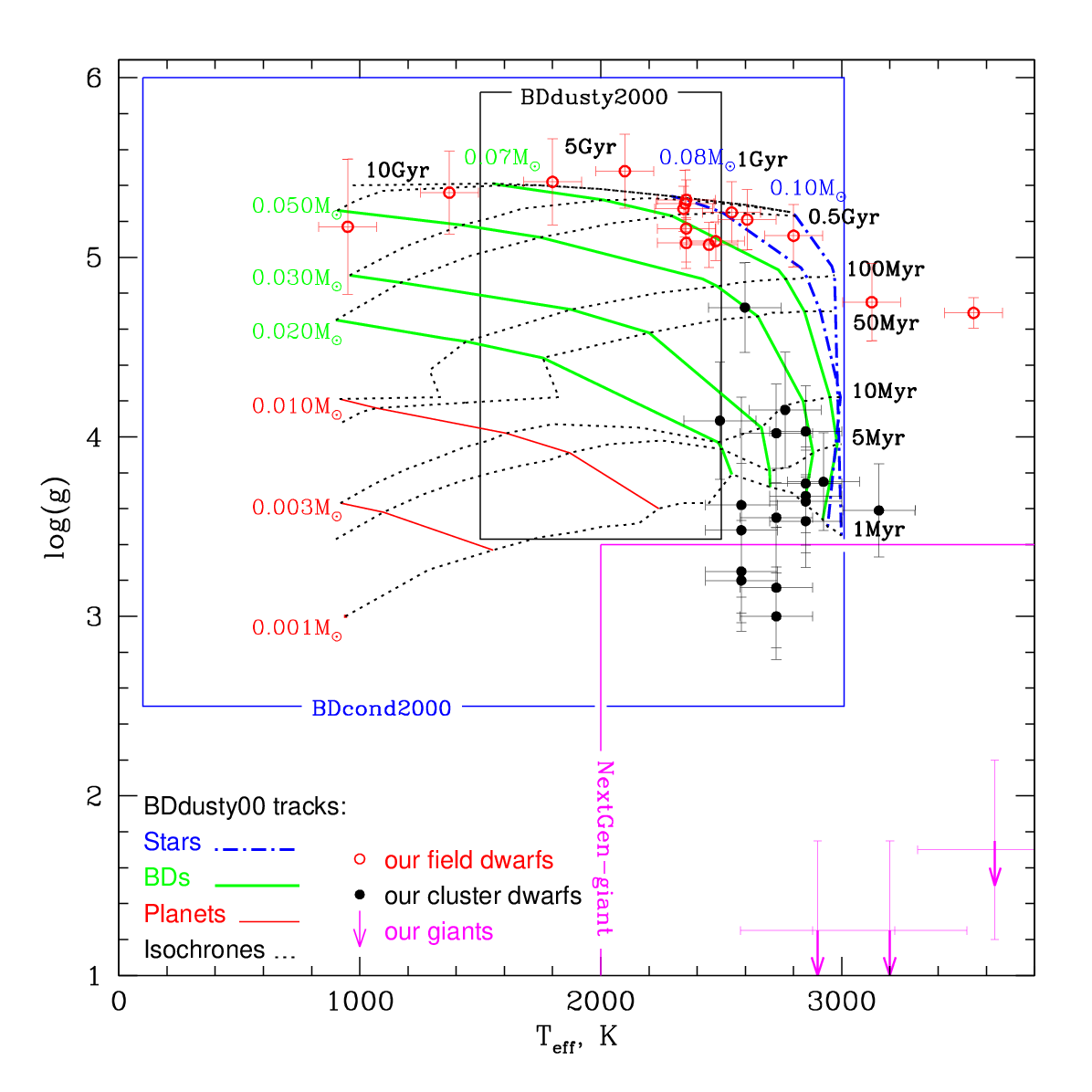}
\caption{\label{hrgT}The \textit{evolutionary models} of \citet{chabrier00} in terms of temperature and gravity. 
Members of young clusters are nicely separated by their gravities
from older dwarfs and from giants of the same temperature. Also shown
as rectangles are the boundaries of \textit{synthetic spectra} grids by the PHOENIX group,
which we used to compare
to our dwarf spectra (BDdusty00 and BDcond00), and to giant HD113285 (NextGen-giant).
}
\end{figure}

Measurement errors also cause uncertainty in the estimated luminosities. Thus, the measurement
errors in m$_J$ (0.05 for the field, 0.1 for cluster members) and the mismatches of the various photometric
systems employed ($\le$ 0.1 magnitudes for 2MASS, CIT, and UKIRT for cool stars
\citep{Carpenter01}) have been combined quadratically to obtain an estimate of $\sigma(m_J)$ = 0.10
magnitudes for field stars and 0.15 for cluster members. In addition, there are errors in the extinction
as a result of uncertainties in both the color measurements and the intrinsic colors. We have
estimated these effects from the relations between SpT and the $I - J$ and $J - K$ colors from
\citet{Dahn02}. We obtain $\sigma (A_J)$ = 0.3 magnitudes (or $\sim$ 1 mag in A$_V$) for cluster objects, and neglect
extinction error for those in the field. In converting m$_J$ to luminosity, we have taken
an uncertainty in BC$_J$ of 0.05 magnitudes for M dwarfs and 0.20 for L dwarfs (based on the
scatter in the BC$_J$ vs. SpT relations in \citet{Dahn02, Reid01, WGM99}).
Finally, we have calculated errors in luminosity due
to uncertainties in the distance moduli individually for each object. For the field dwarfs 
where we have used spectroscopic parallax to estimate the distance moduli, 
we estimate the error to be 0.35 magnitudes.
The resulting total errors in luminosity from combining all of these effects are shown in Figure~\ref{hrD}
and entered in Table~\ref{thisw}. They average 0.18 dex for cluster objects and 0.11 dex for those in the field.

These errors in observational parameters cause uncertainties in the derived parameters such as
mass and gravity. From mass estimates using the three temperature scales and three
sets of theoretical tracks, we find a dispersion for cluster members of 0.024 M$_\odot$, and of
0.012 M$_\odot$ for field stars. An observational contribution to the uncertainty in mass arises
because of the uncertainties in L and T$_{eff}$. We estimate it for each object
from Fig.~\ref{hrD} as the range of masses on evolutionary tracks that pass 
through 1$\sigma$ errorbars in T$_{eff}$ and L. We find
this uncertainty to be on average about 0.025 M$_\odot$ for cluster members and 0.019 M$_\odot$
for field stars. Similarly, the uncertainties in log(g) due to the theoretical tracks and
temperature estimates were estimated from the dispersion in results for individual stars to be 
0.26 dex for cluster members and 0.10 dex in the field. Observational errors in log(g) result
from errors in mass, temperature, and luminosity. Propagating the error estimates
for these quantities, the log(g) errors are 0.30 dex in clusters and 0.18
dex in the field. For both mass and log(g), the observational component is larger and
has been entered in Table~\ref{thisw}.  

To summarize, sources of random error appear to be as large or larger 
than the sources of systematic error considered here.  In all cases, 
we have adopted the ``observational'' random errors in our quantitative 
analysis. 

\subsection{Properties of Giants}\label{gts}

\subsubsection{Masses and Gravities}

For giants, we estimated log(g) from the literature, based on luminosity class and spectral 
type: log(g) = 0.5 $\pm$ 0.5 for M1 -- M8 III stars \citep{Tsuji86,Houdashelt00}. 
The assumed masses of these stars lie between 1 and 5 M$_\odot$. 
The dominant source of uncertainty is variability in SpT and magnitude. Full infrared 
light curves are generally unavailable for long-period variables. Based
on those reported by \citet{Lockwood71}, \citet{Nadzhip01}, and \citet{Bedding02},
we estimate typical variations for our stars to be less than 1 magnitude. No infrared
photometry is available for BD+14 2020, so we estimated its m$_J$ from m$_V$ and the $V - J$
colors taken from \citet{Ducati01} and \citet{Perrin98}. 

Fortunately, the properties of the giants are sufficiently well removed from dwarfs 
(particularly for gravity) that rough estimates suffice for
our purposes.

\subsubsection{Temperatures}
For SpTs between M0 and M7, the giant temperature scales in the literature agree to within $\pm$ 100K,
but by M8 the differences increase to 200K \citep{Perrin98, vanBelle99, ngb}. We used the median of the range of quoted 
SpTs for the giants to determine T$_{eff}$ and BC$_J$ from 
Table 4 of \citet{ngb}. Taking the uncertainty in the temperature calibrations
to be $\sim$ 100K, and an uncertainty in SpT of $\pm$ 2 subtypes, we derive an overall uncertainty in
T$_{eff}$ of 320K. 

\section{Analysis of Near-Infrared Spectral Indices}

Armed with estimates of the temperature and gravity for all objects 
in our sample, along with associated uncertainties, we can now investigate
the correlations of near-infrared spectral features with physical 
properties.

\begin{figure}
\epsscale{1.0}
\plotone{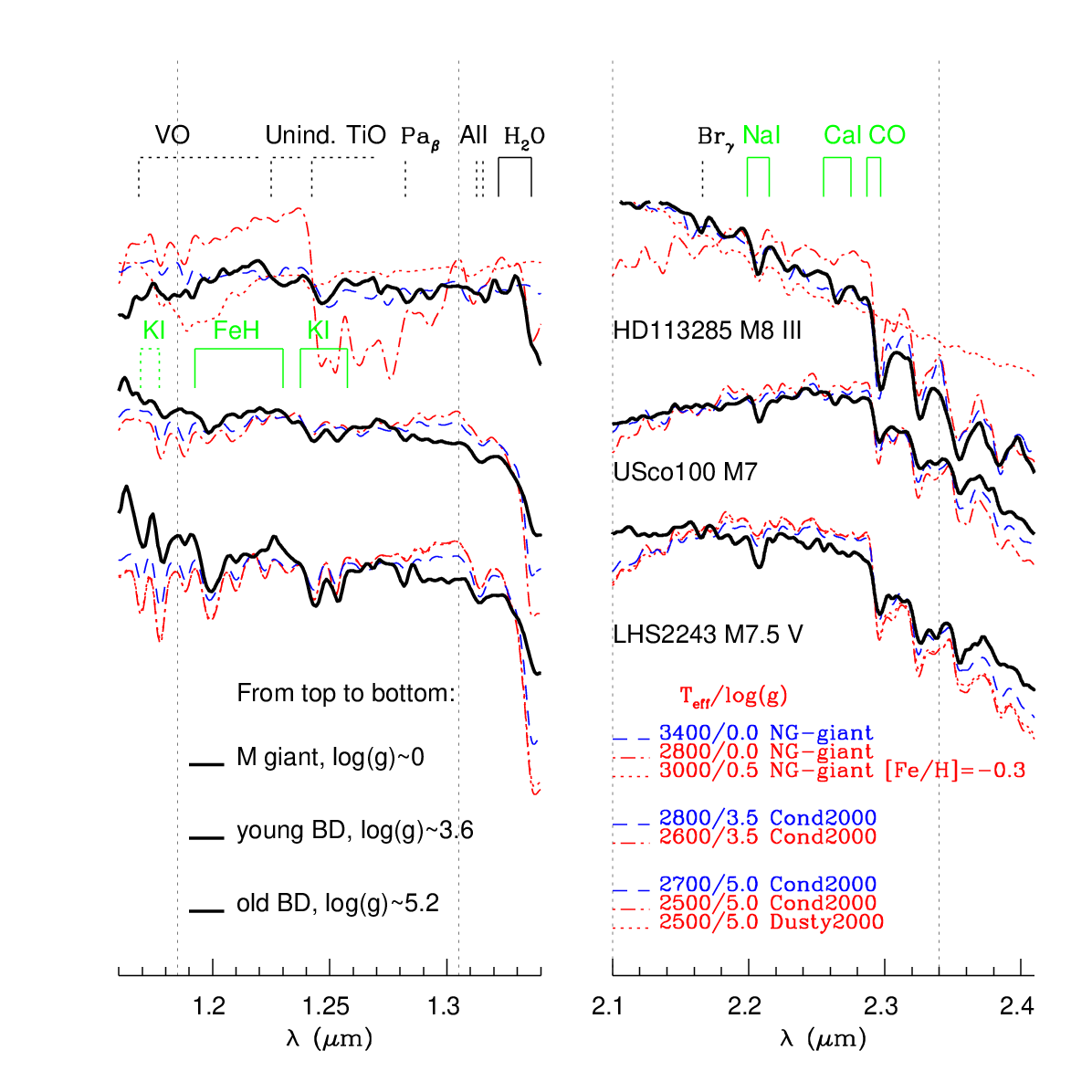}
\caption{\label{mvso}J and K spectra of a giant and two dwarfs of similar SpTs but different ages,
overlaid on the theoretical spectra (of solar metallicity except where noted) 
with the expected range of temperatures and gravities. 
Solid lines define indices analyzed in this work for T$_{eff}$ and gravity sensitivity, 
dotted -- other prominent features including those relevant to giants only. 
\ion{K}{1}, FeH \& \ion{Na}{1} lines are weaker in the younger, larger BD (USco100),
as predicted by the models with lower gravity. 
H$_{2}$O is too strong while \ion{Na}{1} line is too weak in models. 
Except for the H$_{2}$O band at 1.34 $\mu$m, the M giant spectrum is very distinctive with
strong \ion{Na}{1}, \ion{Ca}{1}, CO and the presence of VO and TiO.
Hydrogen lines and the edges of the spectral regions may contain some residuals
from the telluric correction (\S~\ref{datred}). 
The spectra were normalized as described in \S~\ref{datred} -- to have same integrated flux within
the wavelength regions free from strong telluric water absorption (dotted vertical lines),
and offset-ed by constant amount in the vertical direction for clarity. 
}
\end{figure}

In Figure~\ref{mvso} we show candidate spectral indices for temperature and gravity: FeH (1.20$\mu$m), 
\ion{K}{1} (1.25$\mu$m), H$_2$O (1.34$\mu$m), \ion{Na}{1} (2.21$\mu$m), 
\ion{Ca}{1} (2.26$\mu$m), and CO (2.30$\mu$m). 
The indices are defined in Table~\ref{ind}.  They have been selected to measure features 
that are prominent at low spectral resolution \citep{Joyce98, Luhman98a,
McLean00, Wallace00}. The width for H$_2$O and CO (2.30$\mu$m) bands can be
extended depending on the local telluric water vapor content and the desired S/N.
In Table~\ref{indstr}, we report equivalent widths for the  
absorption lines due to FeH, \ion{K}{1}, \ion{Na}{1}, and \ion{Ca}{1} as well as flux ratios 
for H$_2$O and CO. 

First we assess the measurement uncertainties in indices, then the model spectra used for reference, 
and finally analyze the observed dependence of spectral features on temperature and gravity. 

\subsection{Uncertainties in Indices}\label{errewfr}

The most straightforward method to estimate uncertainty in the strength of
an index would be to compare the index strength on a few independent spectra
of a given object. However, due to the reduction process that we adopted
(the spectrum was extracted from a combined 2-dimensional frame,
\textit{not} by combining separately extracted 1-dimensional spectra),
we could not realize this method for all our objects.
Instead, we derived a simple analytic expression that relates an easily measured
S/N in flux (ideally on the featureless part of the combined spectrum) 
to the desired uncertainty in the index strength:

\begin{equation}
\sigma_{index}=\frac{A_{index}}{S/N_{cont}}
\label{sew}
\end{equation}

\noindent
Here, A$_{index}$ represents six empirical coefficients corresponding to our six indices. 
These coefficients were treated as constants and 
were calibrated by inverting Equation~(\ref{sew}) for two of our objects adopted 
as ``error standards'' - 2MASS1707 (field dwarf) and USco67 (cluster member). $\sigma_{index}$ 
in the ``error standards'' was calculated by comparing index values measured in spectra 
extracted from individual, non-combined frames, and by dividing the mean square deviation 
of these values by the square root of the number of frames -- as described at the beginning 
of this section. 

We computed S/N$_{cont}$ for each object as a mean square deviation in flux over the regions
1.2890 - 1.3080$\mu$m (J band) and 2.2230 - 2.2822$\mu$m (K band). 
In these late-type stars, it is impossible to find a 
spectral region free of lines. Hence, S/N$_{cont}$ is only meaningful in a relative sense, 
as an estimate of the quality of the object spectrum relative to that of the ``standards''. 
It is a lower limit on the real S/N in the continuum for giants, where the repeatability 
of features (Figure~\ref{ksp}) indicates the presence of many real lines in these wavelength regions. Then we used expression~(\ref{sew}) to calculate the errors in the index strength for a given star (Table~\ref{indstr}).

The uncertainty associated with the correction for telluric features is not accounted 
in this approach. To assess this effect, one would need to obtain several spectra of 
an object at different times, airmasses, and reduced with different telluric standards. 
Since we were unable to carry out such an extensive study, we are unable to tell 
whether the observed scatter in spectral indices 
at a given SpT and log(g) is due to other relevant 
physical parameters, or to observational errors.  

\subsection{Theoretical Spectra}\label{thsp}

Figure 6 shows the domains of the synthetic spectra calculated by the PHOENIX code of the
Lyon group.\footnote{The model calculations can be found
at http://phoenix.physast.uga.edu} We used the BD Dusty and BD Cond 2000 models \citep{Allard01} to compare to
our observations. These models represent two extreme cases for the treatment of dust. 
The first assumes a uniform distribution of interstellar-sized particles over the whole atmosphere,
with a high level of thermal emission by the dust. The second assumes that the only
important effect of the dust is to remove the refractory elements from the gas (the condensation,
or rain-out model). Emission by dust effectively reduces the equivalent widths of the absorption spectral
features, and it is thought to be relevant in M6 -- L8 dwarfs, while the Cond models should
be applicable to later spectral types. However, we only find significant differences in the 
near infrared between Cond and Dusty model spectra for T$_{eff}$ below 2300K (Figures~\ref{mvso}, \ref{ewg}),
which corresponds to $\sim$ L0, while most of the objects in our sample are of SpT M6 -- M8.
The hotter stars are in the dust-free regime where both models produce identical
results.
 
NextGen models do not consider dust formation and in addition use outdated H$_2$O and TiO
opacities. However, they are the most recent PHOENIX models available for giant gravities, so
we use them in the comparison with the data on giants.

\subsection{Sensitivity to Temperature and Gravity}

\begin{figure}
\epsscale{1.0}
\plotone{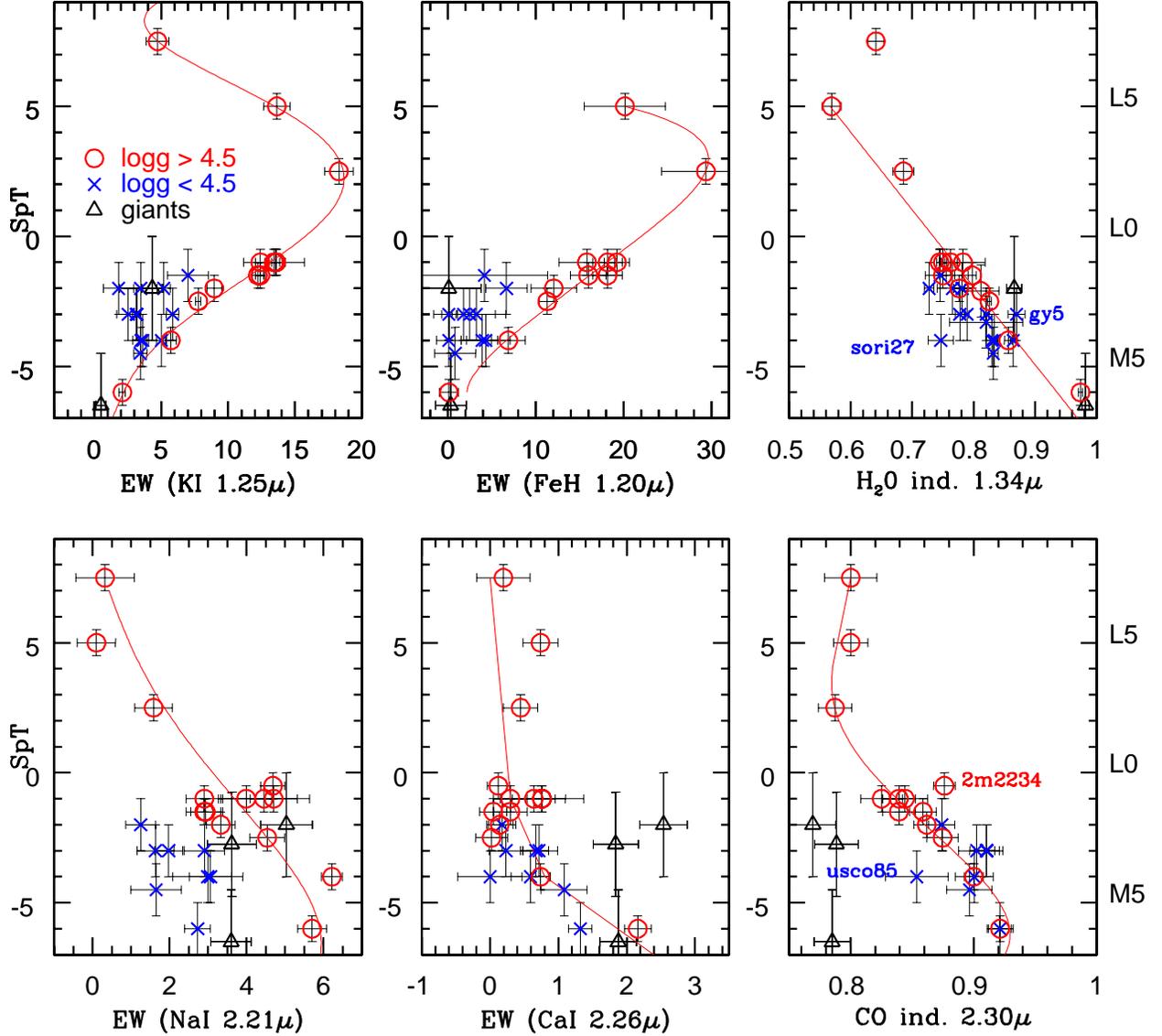}
\caption{\label{ewT}Plot of our index strength against SpT (M5 to L5). Field dwarfs and Pleiades member Pl~17
are represented by circles, cluster members by crosses, and giants by triangles.
One can see that field objects always form an envelope, consistent with them having
larger gravities and narrower gravity range. The scatter is relatively small for H$_{2}$O and CO
indices. Giants occupy a distinctive place on these diagrams. The outliers are identified by name as they
normally have lower S/N. Lines are drawn to represent schematically 
the behavior of the indices for dwarfs.}
\end{figure}

Figure~\ref{ewT}, a plot of SpT vs. index strength, is a traditional way of assigning spectral types.
If we concentrate on {\it field} objects only (circles), we see that most features follow a
smooth sequence, corresponding to change in temperature. For example, the EW of \ion{K}{1} peaks at
L2.5, \ion{Na}{1} at M6, \ion{Ca}{1} earlier than M3, and FeH at L2.5. The broad H$_2$O and CO bands increase 
monotonically in strength to saturation in the mid-Ls. These results are in good agreement
with the higher-resolution near-infrared studies by \citet{Ali95}, \citet{Luhman98a},
\citet{McLean00}, and \citet{Reid01}, among others. They are also consistent with
optical studies of FeH \citep[e.g.][]{Kirkpatrick99, Martin99}. The 
temperatures of peak intensity for alkali-metal lines correlate with their first ionization
potentials. The decline at lower temperatures is due to grain formation and complex
molecular chemistry rather than excitation considerations. 

However, including {\it cluster} objects and {\it giants} (crosses and triangles) introduces
both systematic effects and increased scatter. To explore the possibility that gravity is responsible
for the scatter, we plotted the indices vs. gravity in Figure~\ref{ewg}. We grouped our
objects according to SpT (bearing in mind that the uncertainty in SpTs is $\sim$
1 subtype) and denoted each group by a unique symbol. We also over-plotted measurements from
the theoretical spectra, choosing temperatures representative of our SpT groups.
We discuss the results by specific index.

\subsubsection{\ion{K}{1} \& \ion{Na}{1}}

\ion{K}{1} and \ion{Na}{1} lines in the optical and near-infrared are known to be systematically weaker
in cluster BDs compared to the same SpT in the field \citep[e.g.,][]{Martin96, Luhman98a,
Bejar99, Lucas01, Martin01, Gizis02b}. However, no systematic study of the gravity-dependence is 
available, particularly of the 1.25 $\mu$m feature of \ion{K}{1}, 
which is conveniently situated near the middle of the J band.

As indicated both by the models and our observations, for a given T$_{eff}$ (or SpT), the \ion{K}{1} line
becomes systematically weaker at low gravities. To increase the statistical
significance, we have fitted this behavior after first combining the data for M6 -- M8 subtypes. We
used a linear least-squares routine that takes both the errors of log(g) and EW into account
\citep{Press92}. We obtained the following relation (working range M6 -- M8, 3.0 $>$ log(g)
$>$ 5.5):

\begin{equation}
log(g) = 2.12(\pm0.27) + 0.40(\pm0.05)\times EW(KI_{1.25}, A)
\label{gform}
\end{equation} 

\begin{figure}
\epsscale{1.0}
\plotone{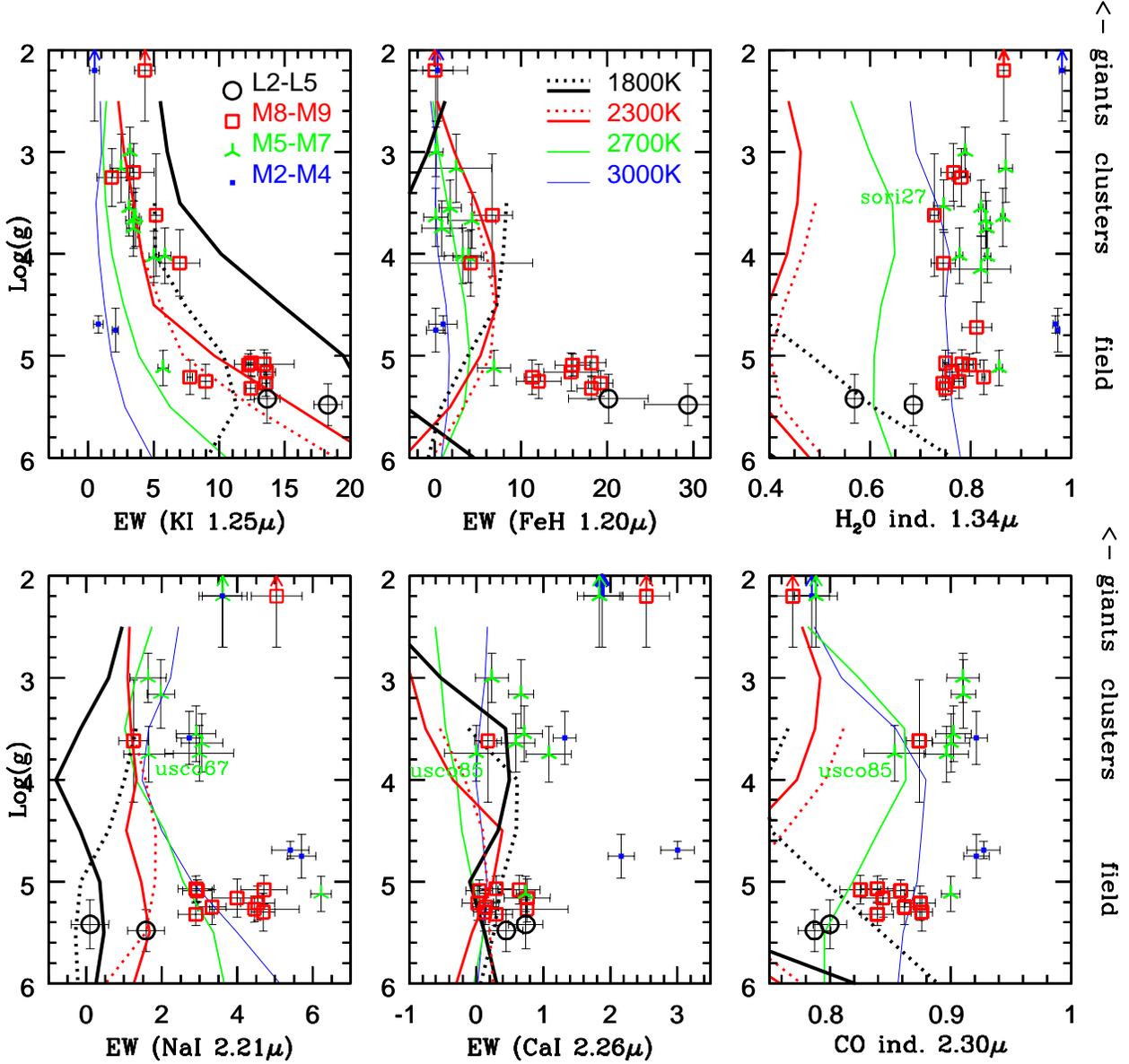}
\caption{\label{ewg}Plots of indices against log(g), as measured in program objects (symbols)
and in models (dotted and dashed lines -- BDdusty00, solid -- BDcond00). Giants are represented
by arrows as their gravities are out of range in this figure ($\sim$0 dex).
Trends with T$_{eff}$ and log(g) are clearly seen for all indices. 
The models generally predict correctly the qualitative behavior of features, but systematic offsets
with observations persist.
}
\end{figure}

Substituting into this expression average values of EW for the cluster
and field dwarfs (3.45 $\pm$ 0.5 \AA, log(g) = 3.5 and 7.71 $\pm$ 0.5 \AA,
log(g) = 5.2, respectively) and quadratically propagating the errors, we find that the
uncertainty in the spectroscopically determined log(g) from
Equation~\ref{gform} is 0.38 dex and 0.51 dex respectively for our low- and high-gravity objects.
The uncertainties are larger than the uncertainty of the "physical" log(g) obtained
from the evolutionary tracks for our objects ($\le$ 0.3 dex, see Section~\ref{err}). This
difference is a natural outcome of the fact that the temperature dependence of these
indices cannot be neglected especially for high gravity objects (see Figures~\ref{ewT}, \ref{ewg}). 
The \ion{Na}{1} line at 2.2 $\mu$m shows a similar trend with log(g) (Figure~\ref{ewg}) but the effect
is weaker. Better modeling of this line is needed as well (Figure~\ref{mvso}).

With more objects, one could construct a more reliable EW -- log(g) -- SpT relation,
especially from intermediate-age clusters (like $\alpha$ Per) where one could probe
the gravity range 4.5 -- 5.0 dex. 

\subsubsection{FeH}

As with the atomic lines just discussed, this molecular feature is both temperature and
gravity sensitive, growing in strength with lower temperatures until the early Ls, and also
with higher gravities. Although this trend is clear, it is difficult to derive a quantitative
relation from our observations, or from models. Observationally, the EWs are uncertain, especially
in cluster objects (where the feature is weak), because the feature is situated at
wavelengths where our S/N is low. Current models fail to reproduce correctly the strong
FeH band in early L dwarfs (see also \citet{McLean00,Leggett01}. For example, 
the Cond and Dusty model predictions are very different for this feature at 1800K as
shown in Figure~\ref{ewg}. There are two reasons for these problems: 1.) the feature is 
very broad and may be contaminated by unidentified lines;
and 2.) the transition from M to L types is marked by the onset of dust formation, which
strongly modifies the pseudo-continuum formed by the adjacent water band. Thus, FeH has
potential as a gravity indicator, but additional observational and theoretical work
is needed to establish its behavior. 

\subsubsection{CO}

The simple interpretation is that gravity is responsible for the difference between the strong
2.3$\mu$m CO band in K -- M giants and the weaker one in dwarfs \citep{Kleinmann86}. This argument
suggests that the strength should be intermediate in young BDs. The smaller ratio of
EW(\ion{Na}{1} + \ion{K}{1})/EW(CO) measured in low-mass cluster members \citep{Greene95, Greene97, Luhman00} compared to the field seems to support this idea. However,
in the absence of reliable SpTs, it is difficult to tell whether the effect was due to 
weakening of \ion{Na}{1} and \ion{Ca}{1}, or to strengthening of CO with lower gravity in late-type stars. 
Thus, \citet{Martin01}
had to invoke veiling from a circumstellar disk when they found that CO was actually weaker in
the Taurus BDs compared to a field dwarf. 

Except for USco85, for which the continuum in our measurements clearly looks problematic at
the red edge (Figure~\ref{ksp}), the CO band does not look much different between cluster and field
objects, in agreement with the results of \citet{Luhman98a} on IC 348 and \citet{Greene95} 
on $\rho$ Oph. Interestingly, the models predict CO to weaken slightly up to log(g) $\le$ 3.8,
without the need to invoke veiling in very cool objects. The models predict 
CO absorption deeper (smaller index values in Figure~\ref{ewg}) 
than that observed, but the agreement in behavior is reasonable for the Cond models. 
Thus, the previously assumed monotonic behavior of CO with gravity does not hold over the range 
of spectral types considered here. The CO lines form a nice monotonic sequence over SpTs
M5 -- L2 for dwarfs, and since they seem to be only slightly gravity-dependent, they can
be used for reliable temperature estimation in young low mass objects. 

\subsubsection{H$_2$O}

The H$_2$O bands are the strongest features in the near-infrared spectra of late M to early L
dwarfs and therefore have been widely used for spectral classification at low resolution \citep[e.g.][]{Testi01}. Detailed comparisons of models and data 
by \citet{Jones95} showed that the water bands are much more
sensitive to temperature than gravity and metallicity. Nevertheless, gravity effects have
been reported in the literature.  Yet, as with CO, the size of these effects remains to be
explored. \citet{Lucas01} claim that for an optically classified young M dwarf
one would expect water absorption in the H band to be as strong as in field L dwarfs. 
However, \citet{Wilking03} find no significant difference in the water absorption strength 
in the K band between M dwarfs and sub-giants.

Our measurements of the water index show a good correlation with SpT. After CO, it
has the least scatter compared with the rest of the indices. There is, however, an indication
that on average, the feature may be stronger in the young objects (Figure~\ref{ewg}). To estimate
this trend quantitatively, we carried our linear fits separately for the low and high gravity
groups (the numbering convention for SpTs is following: M5 $=$ --5, L0 $=$ 0) :

\begin{displaymath}
SpT = 10.99(\pm6.06) - 17.45(\pm7.68)\times(F_{1.34}/F_{1.32}), log(g) < 4.5 
\end{displaymath}
\begin{equation}
SpT = 21.23(\pm7.82) - 29.11(\pm9.56)\times(F_{1.34}/F_{1.32}), log(g) > 4.5 
\end{equation}

\noindent
Since our sample is biased against late-type cluster dwarfs (they are too faint), we are
limited to the SpT range M5.5 -- M8.5 only. The coefficients of the fits differ by $\sim$
1.5 $\sigma$. Thus, use of this index may lead to a slight overestimation of a young BD
SpT in comparison with field dwarfs. However, the effect is within the uncertainty of the
input SpTs of our sample, $\pm$ 1 subtype. Therefore, we confirm the result by \citet{Wilking03}
that, at least within the M spectral types, the water bands can be safely
used (with 1 subtype precision) for the estimation of SpT in low-mass objects, independent
of gravity (age). 

The models confirm this observational conclusion by showing a very weak dependence on log(g),
which in addition is not monotonic. The strengthening of H$_2$O absorption 
at low gravities, which might have been
observed in H band by \citet{Lucas01}, is predicted only for temperatures $\le$ 2000K,
characteristic of L dwarfs. However, even allowing for the uncertainty in the temperature
of $\pm$ 150K, Cond models clearly over-predict the strength of this water band. Dusty
models work better, but only for the high-gravity L dwarfs (Figure 9). 

\subsubsection{\ion{Ca}{1}}

The \ion{Ca}{1} line is strongly temperature sensitive. Unfortunately, it disappears around M7, making 
its measurement uncertain in our objects except for the earliest ones. The gravity sensitivity
of this feature remains to be investigated.

\subsubsection{Giants}

M giants are most easily recognizable by their very deep CO bands, comparable to the ones
in L dwarfs, but they lack the strong water absorption of the latter objects. The H$_2$O
absorption is at least as weak as that in M dwarfs, confirming the recent result of
\citet{Wilking03}. \ion{Na}{1} and \ion{Ca}{1} lines are very prominent, which is the opposite of
what one would expect if they monotonically decreased with gravity. To summarize, giant
spectra are very different from their sub-giant and dwarf star counterparts.  However, 
they do not define an extreme endpoint of a monotonic sequence in observable 
surface gravity effects in the J- and K-band spectra 
over the range of spectral types considered here (M1 -- M8). 

\section{Discussion}

Through our comparison of observed J- and K-band spectra as a function
of surface gravity, and with synthetic spectra from the Lyon group, we have
seen that: 1) CO and H$_2$O are suitable temperature indicators for young M dwarfs;
and 2) \ion{K}{1} and FeH can provide rough estimates of log(g) 
given an estimate for the temperature range of the target star.  Are there
any systematic errors that we have neglected in our analysis that could
mimic the results we have obtained by observing the young, late-type objects
in our sample?  We describe two possible sources of systematic error, 
discuss why we think they do not affect our results, and demonstrate how 
our analysis can be used to confirm membership for candidate members 
of young associations. 

\subsection{Optical Indices and Spectral Classification of Young BDs}\label{optsp}

\begin{figure}
\epsscale{1.0}
\plotone{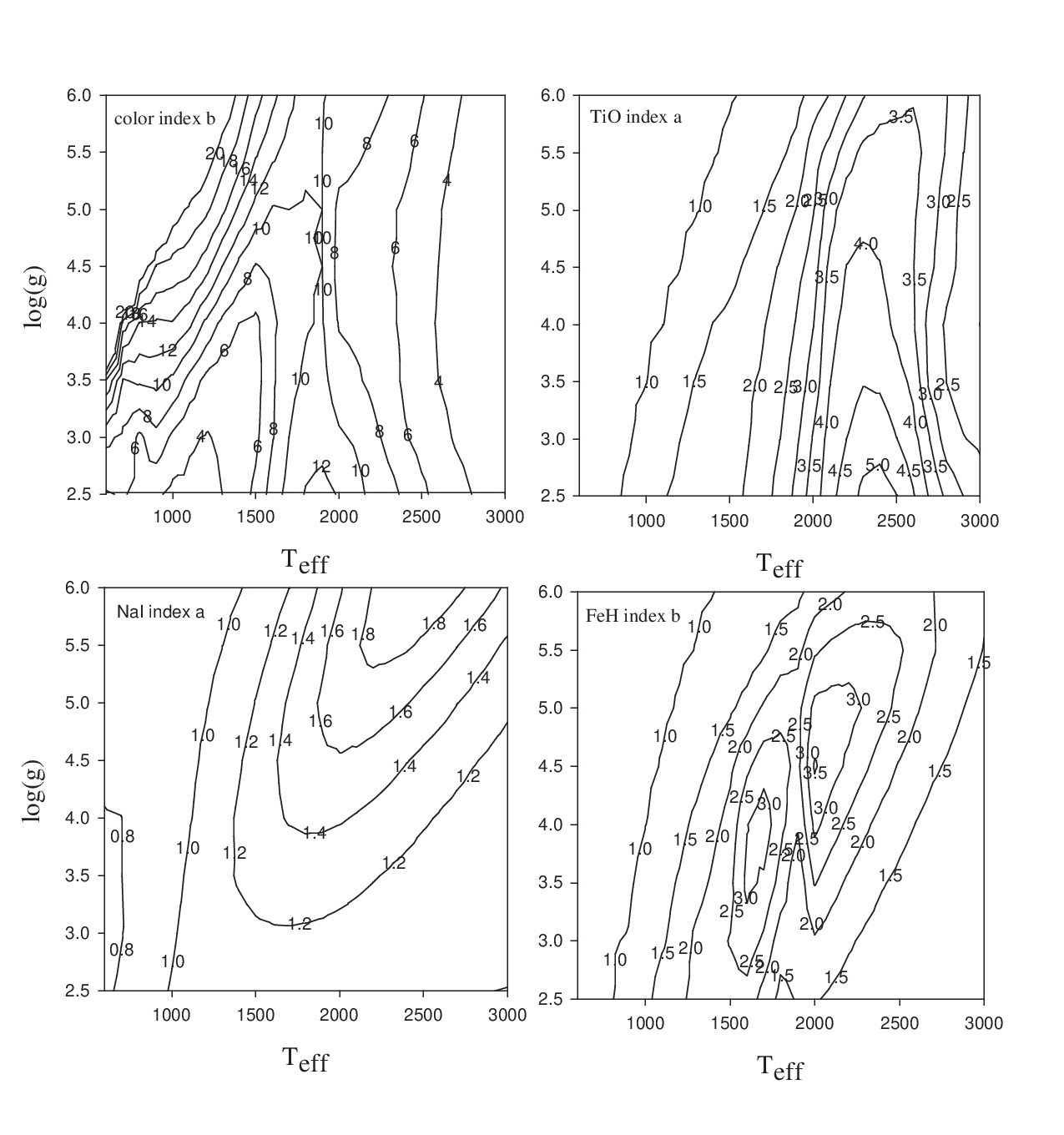}
\caption{\label{optind}The representative contour plots of the indices used for spectral classification
 in the optical, as defined in \citet{Kirkpatrick99}. The indices were
measured on BDcond00 model spectra, smoothed to the resolution of 9 \AA.
The figure shows characteristic behavior of certain species,
repeated also in the IR. One can see that the regions on log(g)--T$_{eff}$ diagram
where a given index would depend on only one parameter out of the two,
are restricted. However, ``color b'' and ``TiO a'' indices are
appropriate for most of the objects in our sample (where contours are nearly straight
vertical lines).
}
\end{figure}

Perhaps the adopted spectral types (obtained 
from visible wavelength spectra) for PMS stars in our sample
are systematically in error as a function of surface gravity. 
We need to determine the gravity-dependence of the indices used for spectral classification
in the {\it optical} (Table 1) to ensure that the existing SpT classification, which is
based on comparison to field dwarfs, can be applied to BDs with sub-giant gravities. 
To investigate this, we measured indices
defined in \citet{Kirkpatrick99} in BDcond00 model spectra (smoothed to the 9 \AA~ resolution
 used by \citet{Kirkpatrick99}) and plotted the contours of equal index
strength on the log(g) -- T$_{eff}$ plane. A few characteristic patterns are shown in Figure~\ref{optind}. 
It is satisfying to see that ``color'' indices and the oxide molecules (TiO, VO, H$_2$O, CO)
that are most frequently used for spectral classification of M stars, are relatively
gravity-independent.

However, this behavior holds only over a limited range of parameter space. For example, for the
spectral classification of sources with log(g) ranging from 3 to 5 dex, one would like to use
indices behaving similarly to the "color-b" index for M5 -- M9 objects ($\sim$ 2300 -- 3000K)
and to use "TiO-a" for M5 -- M7 ($\sim$ 2700 -- 3000K). After estimating the SpT, one could then
use atomic lines (\ion{Na}{1}, \ion{K}{1}) and possible hydride molecular bands (FeH, CrH, CaH) to infer
gravity (and hence young cluster membership) for the objects of interest.
More general lessons are 1) to be aware of strong gravity dependance where
$\frac{\partial index}{\partial T_{eff}} = 0$ (Fig.~\ref{optind}) and 2) always assign spectral
type based on a few different indices -- in order to avoid biases associated with the individual
gravity dependence of a single index.  

However, we prefer calibration of object indices on standard stars, rather than directly to
models. Models can provide the correct qualitative behavior, but not necessarily a good quantitative
match. For example, if one tries to use the water band at 1.3 -- 1.6$\mu$m and PHOENIX spectra,
even with the proper gravity one will overestimate the temperature because the synthetic
band is obviously too deep for a given T$_{eff}$ (our Figures~\ref{mvso}, \ref{ewg}; see also \citet[][Fig.9]{Leggett01}; and \citet[][their Fig.3]{Natta02}). This effect is our 
explanation for the very hot temperatures derived by \citet{Lucas01} for young sources
in Orion.

Uncertainties in current spectral synthesis models include: 
1) incompleteness of the opacity lists \citep{Burrows02, Cushing02}; 
2) the treatment of dust \citep{Allard01,Tsuji02}; 3) the dependence on initial 
conditions for young objects \citep{Baraffe02}, as well as other effects. 
Because of these limitations, we did not try to derive our
own temperature scale by comparing J and K spectra with synthetic spectra. Some of the dangers
are exposed by apparent inconsistencies; for example, from Figure~\ref{ewg}, one would conclude that 
M6 corresponds to T$_{eff}$ $>$ 3000K looking at the \ion{Na}{1}, H$_2$O, and CO features, while
according to \ion{K}{1} and FeH, M6 corresponds roughly to 2500K.

\subsection{Other Effects at Young Ages}

Perhaps our spectra are affected by activity associated with youth that
could mimic the observed effects we attribute to surface gravity. 
One could argue that effects such as continuum veiling from
accreting material could make lines weaker in cluster objects. 
For example, \citet{Jayawardhana02} have discovered a broad, asymmetric H$\alpha$ line in GY5, indicative
of on-going accretion. \citet{Martin01} report emission in Br$\gamma$ in CFHT-BD-Tau4,
as well as strong H$\alpha$. These observations 
extend into the BD range the well-known
tendency for activity associated with youth observed in low mass PMS stars. 
Such accretion activity tends to be accompanied by continuum 
excess emission in the near-infrared that can dilute photospheric features
\citep{Hartigan95}. 

According to our data, both GY5 and CFHT-BD-Tau4, members of the 0.3--1 Myr old embedded $\rho$ Oph 
and 1--3 Myr Taurus-Auriga T association respectively, have weaker \ion{Na}{1} and CO absorptions
compared to the similar SpT USco100. 
USco100, an M7 member of the 5 Myr old Upper Sco subgroup of the Sco Cen OB association, 
has normal, single-peaked, H$\alpha$ emission, characteristic of coronal rather than disk accretion
origin \citep{Jayawardhana02}, but so does GY5 according to \citet{Muzerolle03}. So it is
more plausible that the difference in absorption line strength that we are detecting 
is due to the expected age difference, reflected in lower log(g) for
GY5 and CFHT-BD-Tau4. The evidence against veiling as an explanation is four-fold.
First, little veiling is expected for the very low mass stars given the low
mass accretion rates from their disks, as discussed by \citet{Muzerolle03}.
Second, veiling should affect both molecular and atomic lines, while it is only \ion{Na}{1} and \ion{K}{1},
shown here to be strongly gravity-dependent, that show significant differences between
cluster members and field objects. Third, neither GY5 nor CFHT-BD-Tau4 show 
significant evidence for continuum excess in their broad-band colors.  
Finally, even if moderate veiling is 
influencing the index strengths in the K band, it should be much less important in the 
J band \citep[][Fig.2]{Meyer97}. 
It is unlikely that veiling is responsible for the factors of
2 -- 3 difference in the line strengths between cluster and field dwarfs seen in the J band.
In any case, both veiling and gravity work in the
same direction, to decrease the line strengths in young objects. Therefore, weak 
atomic lines can be used for reliable identification of members in young 
clusters, regardless of the mechanism by which they are weakened. 

\subsection{Examples}

A particularly valuable application of the gravity-sensitive indices is to identify
members of nearby associations, which are typically spread over large areas on the
sky and hence are seriously contaminated by interlopers. We illustrate this
technique with two examples. 
2MASS1139-3159 was identified as an M8 member of the young TW Hya association, based on the
strength of VO and the weakness of CaH and \ion{Na}{1} in the optical \citep{Gizis02b}. The weakness
of the FeH and \ion{K}{1} lines in our J-band spectrum, and of the \ion{Na}{1} line in the K spectrum compared
to the field M8 dwarf 2MASS1444+3002 confirms the low gravity of this object (see Figures~\ref{jsp}, \ref{ksp}). 
The same effect is observed in our spectra of one of the Taurus-Auriga 
association members, CFHT-BD-Tau4 (M7).
Its membership was established by the weak \ion{K}{1} and \ion{Na}{1} lines in the optical and K bands
\citep{Martin01}, and we confirm this behavior in the K and J bands. 
As demonstrated above, spectrophotometry with SNR $\sim$ 50--100 derived
from narrow-band filters of R $\sim$ 50--100, can be quite useful in estimating 
temperatures for very cool stars and distinguishing both foreground and background 
stars from bona fide cluster members.  Additional work is needed to define
a set of optimum filters at the lowest possible spectral resolution to 
enable efficient discrimination at modest sensitivity. 

\section{Summary}

By inter-comparison of low resolution (R $\sim$ 300 -- 350) J- and K-band spectra of field and
young cluster M dwarfs and M giants, and with guidance from recent model atmosphere
calculations, we have arrived at the following conclusions regarding the temperature
and gravity sensitivity of a number of strong near-infrared spectral features:

\begin{itemize}

\item
The H$_2$O and CO bands at 1.35 and 2.30$\mu$m are relatively insensitive to gravity
in the spectral range of M dwarfs, and therefore can be used for spectral classification
on young stellar objects (with $\sim$ 1 subtype precision). 

\item
Alkali-metal lines of \ion{K}{1} and \ion{Na}{1} at 1.25 and 2.21$\mu$m (and possible FeH at 1.20$\mu$m)
can be used to derive log(g) with an uncertainty of $\pm$ 0.5 dex from R $\sim$ 350,
S/N $\ge$ 70 spectra of M4 -- M9, 3.0 $>$ log(g) $>$ 5.5 objects, given that the spectral
type is known to $\pm$ 1 subtype. 

\item
The spectra of late M giants are very distinctive from those of substellar objects. 
These differences can be explained in terms of their extremely low surface gravities,
hotter temperature scale, and non-equilibrium processes in their tenuous atmospheres.

\item
The current models are unable to reproduce accurately all the observed features in
these stars, and therefore should be used with caution for determination of physical
parameters from spectra, especially from a single line or limited set of lines. 

\end{itemize}

\acknowledgments

We would like to thank Marcia Rieke, Erick Young and John Stansberry
for help with observations; Chad Engelbracht, Valentin Ivanov, Amaya Moro-Martin and Kevin Luhman
for help with data reduction; and Adam Burrows and France Allard for useful discussions.
NG was supported by the scholarship from the Jesuits of the Vatican Observatory;
MM acknowledges support from Lucas Grant.

\clearpage


\clearpage

\begin{deluxetable}{llrrrr}
\tabletypesize{\small}
\tablecaption{Program object properties taken from the literature \label{litr}}
\tablewidth{0pt}
\tablehead{
\colhead{object name} & \colhead{SpT \tablenotemark{a}} & \colhead{apparent mag} &
\colhead{A$_{V}$} & \colhead{distance, pc} & \colhead{references \tablenotemark{b}}} 
\startdata
\textit{Cluster members:} &  & & & & \\
IC348-355 & M8 o   & J $=$ 14.65 & 0.3 & 318 $\pm$ 27 & 1,2,2,3 \\
IC348-363 & M8 o   & J $=$ 14.83 & 0.0 & 318 $\pm$ 27 &  2,2,2,3 \\
CFHT-BD-Tau2 & M8 o   & J $=$ 13.76 & 0.0 & 140 $\pm$ 10 &  4,4,4,5 \\
CFHT-BD-Tau4 & M7 o   & J $=$ 12.16 & 3.0 & 140 $\pm$ 10 & 4,4,4,5 \\
SOri 27 & M6 o & J $=$ 14.86 & 0.0 & 473 $\pm$ 33 &  6,6,6,3 \\
CRBR 31   & M6.7 nir & J $=$ 16.57 & 8.6 & 125 $\pm$ 25 & 7,8,7,3 \\
GY 5 (CRBR 21) & M7 nir & J $=$ 12.70 & 4.5 & 125 $\pm$ 25 & 10,8,10,3 \\
GY 141    & M8.5 o & J $=$ 15.13 & 0.0 & 125 $\pm$ 25 & 11,8,11,3 \\
WL 14 (CRBR 52, GY 172) & M4 nir & J $=$ 16.12 & 19.2 & 125 $\pm$ 25 & 12,8,12,3 \\
2MASSW J1139511-315921 & M8 o & J $=$ 12.67 & 0.0 & $73 ^{+25}_{-55}$ &  13,13,,14 \\
USco 66   & M6 o   & I $=$ 14.85 & 0.0  & 145 $\pm$ 2  &  9,9,9,3 \\
USco 67   & M5.5 o & I $=$ 14.87 & 0.5  & 145 $\pm$ 2  & 9,9,9,3 \\  
USco 75   & M6 o   & I $=$ 15.08 & 0.0  & 145 $\pm$ 2  & 9,9,9,3 \\   
USco 85   & M6 o   & J $=$ 13.03 & 0.0  & 145 $\pm$ 2  & 9,9,9,3 \\  
USco 100  & M7 o   & I $=$ 15.62 & 0.0  & 145 $\pm$ 2  & 9,9,9,3 \\   
USco 109  & M6 o   & I $=$ 16.06 & 0.0  & 145 $\pm$ 2  & 9,9,9,3 \\
USco 128  & M7 o   & J $=$ 14.41 & 0.0  & 145 $\pm$ 2  & 9,9,9,3 \\
CFHT-Pl 17 & M7.9 o & J $=$ 16.01 & 0.1 & 132 $\pm$ 15 & 15,15,15,16 \\
\\
\textit{Field dwarfs:}  & & & & &  \\
GJ 402 & M4e o & J $=$ 7.30 & 0.0 & 5.6 $\pm$ 0.7 & 17,17,,18 \\
 Gl 569A & M0e/M3\tablenotemark{c} o & V $=$ 10.14 & 0.0 & 9.8 $\pm$ 0.2 & 19/20,21,,18 \\
 Gl 569Ba & M8.5 nir & J $=$ 11.14 & 0.0 & 9.8 $\pm$ 0.2 &  22,22,,18 \\
Gl 569Bb &  M9 nir & J $=$ 11.65 & 0.0 & 9.8 $\pm$ 0.2  & 22,22,,18 \\
BRI 1222-1221 & M9 o & J $=$ 12.56 & 0.0 & 17.1 $\pm$ 1 & 23,23,,24 \\
  LHS 2243 & M7.5 o & J $=$ 11.95 & 0.0 & - & 23,23,, \\
 LHS 2397a\tablenotemark{d} & M8.5 o & J $=$ 11.93 & 0.0 & 14.3 $\pm$ 0.4 & 23,23,,35 \\
LHS 2924 & M9 o & J $=$ 11.84 & 0.0 & 11.0 $\pm$ 0.2 & 17,17,,35 \\
2MASSI J0825196+211552 & L7.5 o & J $=$ 15.12 & 0.0 & 10.7 $\pm$ 0.5 & 25,25,,24 \\
2MASSI J1029216+162652 & L2.5 o & J $=$ 14.31 & 0.0 & - & 25,25,, \\
2MASSW J1049414+253852 & M6 o & J $=$ 12.40 & 0.0 & - & 23,23,, \\
2MASSW J1239272+551537\tablenotemark{e} & L5 o & J $=$ 14.67 & 0.0 & - & 25,25,, \\
2MASSW J1444171+300214 & M8 o & J $=$ 11.68 & 0.0 & - & 23,23,, \\
2MASSW J1707183+643933 & M9 o & J $=$ 12.56 & 0.0 & - & 23,23,, \\
2MASSI J2234138+235956 & M9.5 o & J $=$ 13.14 & 0.0 & - & 23,23,, \\
SDSSp 162414.37+002915.6 & T6 nir &  J $=$ 15.53 & 0.0 & 10.9 $\pm$ 0.3 & 26,27,,24 \\
\\
\textit{Giants:}  & & & & & \\
BD+14 2020  & M2/M5\tablenotemark{c}  III o & V $=$ 10.1 & -  & -  & 28/29,30,, \\
HD 113285 (RT Vir) & M1/M8\tablenotemark{f} III o & J $=$ 0.36  $\pm$ 1 & 0.0  & 138 $\pm$ 20 & 28/31,32,,18 \\
HD 126327 (RX Boo) & M6.5e-M8e\tablenotemark{c} III o & J $=$ -0.59 $\pm$ 1 & 0.0  & 156 $\pm$ 24 &31,33,34,18 \\ 
\enddata

\tablenotetext{a}{``o'' means optical SpT, ``nir'' -- near-IR}
\tablenotetext{b}{references given in the following order: reference for SpT,
photometry, extinction, distance}
\tablenotetext{c}{mean SpT was adopted for T$_{eff}$ estimation}
\tablenotetext{d}{M8/L7.5 binary with $\Delta$J=3.83 according to \citet{Freed03}}
\tablenotetext{e}{L5/L5 binary according to \citet{Bouy03}}
\tablenotetext{f}{IR colors favor M8, which was further adopted}
\tablerefs{
(1) Luhman et al. 1998; 
(2) Luhman 1999; 
(3) de Zeeuw et al. 1999;
(4) Mart\'{\i}n et al. 2001;
(5) Kenyon, Dobrzycka, \& Hartmann 1994;
(6) B\'{e}jar et al. 2001;
(7) Cushing, Tokunaga, \& Kobayashi 2000;
(8) Barsony et al. 1997;
(9) Ardila, Mart\'{\i}n, \& Basri 2000;
(10) Wilking, Greene, \& Meyer 1999;
(11) Luhman, Liebert, \& Rieke 1997;
(12) Luhman \& Rieke 1999;
(13) Gizis 2002;
(14) Makarov \& Fabricius 2001; 
(15) Mart\'{\i}n at al. 2000;
(16) Stello \& Nissen 2001;
(17) Leggett 1992;
(18) The Hipparcos Catalog 1997;
(19) The Third Catalog of Nearby Stars 1991;
(20) Montes et al. 1997;
(21) Hawley, Gizis, \& Reid 1996;
(22) Lane et al. 2001;
(23) Gizis et al. 2000;
(24) Dahn et al. 2002;
(25) Kirkpatrick et al. 2000;
(26) Geballe et al. 2002;
(27) Strauss et al. 1999;
(28) AGK3 Catalog 1975;
(29) Jaschek, Conde \& de Sierra 1964;
(30) The Hipparcos and Tycho Catalogs 1997;
(31) GCVS 1998;
(32) Kerschbaum \& Hron 1994;
(33) Ducati 2002;
(34) Perrin et al. 1998;
(35) Monet et al. 1992.
}
\end{deluxetable}   

\begin{deluxetable}{lrlcrrl}
\tablewidth{0pt}
\tabletypesize{\small}
\tablecaption{Observing log \label{obs}} 
\tablehead{
\colhead{object name} & \colhead{mag \tablenotemark{a}}    & \colhead{band} & 
\colhead{Yr-MM-DD}    &  \colhead{t$_{exp}$ \tablenotemark{b}}  &
\colhead{``S/N'' \tablenotemark{c}} & \colhead{Telluric Standard}}
\startdata
\textit{Cluster members:} &  & & & & &  \\
IC348-355 & 14.7 & J & 2001-03-06 & 1920 & 44 & HD 26710 G2V \\ 
IC348-363 & 14.8 & J & 2001-03-06 & 1920 & 62 & HD 23169 G2V \\
CFHT-BD-Tau2 & 12.2 & K & 2002-03-26 & 1440 & 39 & HD 27741 G0V \\
CFHT-BD-Tau4 & 12.2 & J & 2002-03-26 & 960 & 396 & HD 27741 G0V \\
             & 10.3 & K & 2002-03-26 & 480 & 141 & HD 27741 G0V \\
SOri 27 &      14.9 & J & 2001-03-06 & 1920 & 53 & HD 160933 F9V \\
CRBR 31   & 16.6 & J & 2001-05-06 & 2880 & 18 & HD 146997 G2V \\
GY 5 & 12.7 & J & 2002-03-27 & 1320 & 80  & HD 147284 G3V \\ 
 & 10.9 & K & 2002-03-29 & 960 & 185 & HD 147284 G3V \\  
GY 141    & 15.1 & J & 2001-05-07 & 2880 & 46 & HD 146997 G2V \\ 
WL 14 & 16.1 & J & 2001-05-07 & 2880 & 18 &  HD 146997 G2V \\ 
      &  11.7 & K & 2001-05-07 & 960 & 205 &  HD 146997 G2V \\ 
2MASSW J1139511-315921 & 12.7 & J & 2002-03-29 & 960 & 141 & HD 100585 G3V \\
  &           11.5             & K & 2002-03-26 & 480 & 176 & HD 100585 G3V \\
USco 66   & $\approx$ 12.7 & J & 2002-03-27 & 1920 & 233 & HD 141092 G3V \\ 
          & $\approx$ 12.0     & K & 2002-03-27 & 960 & 123 & HD 141092 G3V \\  
USco 67   & $\approx$ 12.7 & J & 2002-03-29 & 2400 & 142 & HD 142523 G1/G2V \\
          & $\approx$ 11.8     & K & 2002-03-29 & 960 & 104 & HD 142523 G1/G2V \\ 
USco 75   & $\approx$ 12.9 & J & 2001-05-06 & 960 & 118 & HD 146997 G2V \\
USco 85   & $\approx$ 12.1 & K &  2002-03-28 & 960 & 75 & HD 142523 G1/G2V \\  
USco 100  & $\approx$ 13.2 & J & 2002-03-27 & 1440 & 253 & HD 141092 G3V \\
          & $\approx$ 12.2  & K & 2002-03-27 & 960 & 130 & HD 141092 G3V \\
USco 109  & $\approx$ 13.9 & J & 2001-05-06 & 960 & 185 & HD 146997 G2V \\ 
USco 128  & 14.4 & J & 2002-03-27 & 2400 & 152 & HD 142523 G1/G2V \\
          & $\approx$ 13.5 & K & 2002-03-29 & 1440 & 47 & HD 142523 G1/G2V \\
CFHT-Pl 17 & 16.0 & J & 2001-03-06 & 3840 & 37 & HD 23169 G2V \\
\\
\textit{Field dwarfs:}  & & & & & &  \\
GJ 402 & 7.3 & J & 2002-03-28 & 160 & 303 & HD 88725 G1V \\
& 6.4  & K & 2002-03-28 & 160 & 180 & HD 88725 G1V \\
Gl 569A & $\approx$ 7.0 & J & 2001-03-04 & 20 & 202 & HD 131473 F9V \\
&  $\approx$ 6.2 & K & 2001-03-04 & 260 & 139 & HD 131473 F9V \\
Gl 569Bab\tablenotemark{d} & 10.6 & J & 2001-03-04 & 960 & 199 & HD 131473 F9V \\  
& 9.5  & K & 2001-03-04 & 260 & 142 & HD 131473 F9V \\  
BRI 1222-1221 & 12.6 & J & 2001-03-05 & 480 & 31 & HD 108799 G1/G2V \\
& 11.4 & K &  2001-03-05 & 240  & 111 & HD 108799 G1/G2V \\
LHS 2243 & 12.0 & J & 2002-03-26 & 1440 & 351 & HD 91950 G2V \\
           & 11.0 & K & 2002-03-26 & 480 & 150 & HD 91950 G2V \\
LHS 2397a & 11.9 & J & 2002-03-27 & 960 & 164 & HD 98298 G3V \\  
           & 10.7 & K & 2002-03-27 & 480 & 176 & HD 98298 G3V \\ 
LHS 2924 & 11.8 & J & 2001-03-05 & 480 & 228 & HD 134044 F8V \\ 
          & 10.7 & K & 2001-03-05 & 480 & 57 & HD 134044 F8V \\  
2MASSI J0825196+211552 & 15.1 & J & 2001-05-06 & 2880 & 83 & HD 68255 F9V \\ 
                       & 13.1 & K & 2001-05-07 & 1440 & 90 & HD 68257 F7V \\
2MASSI J1029216+162652 & 14.3 & J & 2001-05-06 & 960  & 66 & HD 87776 G0V \\
                       & 12.6 & K & 2001-05-06 & 960 & 140 & HD 87776 G0V \\
2MASSW J1049414+253852 & 12.4 & J & 2002-03-26 & 960 & 173 & HD 91950 G2V \\
                       & 11.4 & K & 2002-03-26 & 960 & 250 & HD 91950 G2V \\
2MASSW J1239272+551537 & 14.7 & J & 2001-05-07 & 1920 & 71 & HD 108954 F9V \\
                        & 12.7 & K & 2001-05-08 & 2400 & 137 & HD 108954 F9V \\
2MASSW J1444171+300214 & 11.7 & J & 2001-03-05 & 480 & 127 & HD 129357 G2V \\
                       & 10.6 & K & 2001-03-05 & 240 & 183 & HD 129357 G2V \\
2MASSW J1707183+643933 & 12.6 & J & 2001-03-05 & 480 & 105 & HD 160933 F9V \\
                       & 11.4 & K & 2001-03-05 & 240 & 103 &  HD 160933 F9V \\
2MASSI J2234138+235956 & 11.8 & K & 2001-05-07 & 960 & 216 & HD 210211 G2V \\
SDSSp 162414.37+002915.6 & 15.5 & J & 2001-05-08 & 2880 & 21 & HD 140538 G2.5V \\
\\
\textit{Giants:}  & & & & & & \\
BD+14 2020  & $\approx$ 6 $\pm$ 1 & J & 2002-03-27 & 160 & 188 & HD 75528 G1V \\
            & $\approx$ 5 $\pm$ 1            & K & 2002-03-27 & 160 & 130 & HD 75528 G1V \\
HD 113285 & 0.4 $\pm$ 1 & J & 2002-03-27 & 20 & 89 & HD 114606 G1V \\
                   &  -1.1 $\pm$ 1    & K & 2002-03-28 & 30 & 101 & HD 114606 G1V \\
HD 126327 & -2.0 $\pm$ 1 & K & 2002-03-28 & 100 & 107 & HD 126991 G2V \\ 
\enddata

\tablenotetext{a}{``$\approx$'' means that the given magnitude was estimated from other magnitudes,
which are reported in Table~\ref{litr}}
\tablenotetext{b}{total exposure time in seconds}
\tablenotetext{c}{``S/N'' is the ratio of the average flux and dispersion 
between 1.29$-$1.31 and 2.22$-$2.26 $\mu$m
measured on the final, smoothed spectra.
This is a good measurement of the relative quality of spectra, see \S~\ref{errewfr} for details}
\tablenotetext{d}{composite spectrum}
\end{deluxetable}   

\clearpage

\begin{deluxetable}{lrrrrr}
\tabletypesize{\small}
\tablecaption{Program object properties derived in this work \label{thisw}}
\tablewidth{0pt}
\tablehead{
\colhead{object name} & \colhead{(M$_{J}$)$_{0}$} & \colhead{log(L/L$_{\odot}$)} &
\colhead{T$_{eff}$ \tablenotemark{a}} & 
\colhead{M/M$_{\odot}$ \tablenotemark{a}} & \colhead{log(g) \tablenotemark{a}}}
\startdata
\textit{Cluster members:} &  & & & & \\
IC348-355         &7.06   &   -1.76 $\pm$ 0.15 &   2583 $\pm$ 150 &  0.028 $\pm$ 0.015 &  3.20 $\pm$ 0.28   \\
IC348-363         &7.32   &   -1.87 $\pm$ 0.15 &   2583 $\pm$ 150 &  0.026 $\pm$ 0.015 &  3.25 $\pm$ 0.28   \\
CFHT-BD-Tau2      &8.03   &   -2.15 $\pm$ 0.15 &   2583 $\pm$ 150 &  0.022 $\pm$ 0.015 &  3.48 $\pm$ 0.37   \\
CFHT-BD-Tau4      &5.63  &   -1.21 $\pm$ 0.15 &   2729 $\pm$ 150 &  0.058 $\pm$ 0.030 &  3.00 $\pm$ 0.24   \\
SOri 27           &6.48   &   -1.55 $\pm$ 0.15 &   2851 $\pm$ 150 &  0.069 $\pm$ 0.035 &  3.53 $\pm$ 0.26   \\
CRBR 31            &8.81  &   -2.48 $\pm$ 0.22 &   2765 $\pm$ 150 &  0.041 $\pm$ 0.015 &  4.15 $\pm$ 0.32   \\
GY 5     &6.02  &   -1.37 $\pm$ 0.22 &   2729 $\pm$ 150 &  0.056 $\pm$ 0.035 &  3.16 $\pm$ 0.34   \\
GY 141             &9.64  &   -2.79 $\pm$ 0.22 &   2494 $\pm$ 150 &  0.024 $\pm$ 0.015 &  4.09 $\pm$ 0.33   \\
WL 14     &5.54  &   -1.14 $\pm$ 0.22 &   3155 $\pm$ 150 &  0.144 $\pm$ 0.040 &  3.59 $\pm$ 0.26   \\
2MASSW J1139511-315921     &8.35  &   -2.28 $\pm$ 0.50 &   2583 $\pm$ 150 &  0.022 $\pm$ 0.015 &  3.62 $\pm$ 0.60   \\
USco 66            &6.84   &   -1.69 $\pm$ 0.14 &   2851 $\pm$ 150 &  0.066 $\pm$ 0.035 &  3.64 $\pm$ 0.29   \\
USco 67            &6.61   &   -1.60 $\pm$ 0.14 &   2924 $\pm$ 150 &  0.081 $\pm$ 0.040 &  3.75 $\pm$ 0.27   \\
USco 75            &7.07   &   -1.78 $\pm$ 0.14 &   2851 $\pm$ 150 &  0.059 $\pm$ 0.030 &  3.67 $\pm$ 0.27   \\
USco 85            &7.22   &   -1.84 $\pm$ 0.14 &   2851 $\pm$ 150 &  0.058 $\pm$ 0.030 &  3.74 $\pm$ 0.27   \\
USco 100           &7.41   &   -1.92 $\pm$ 0.14 &   2672 $\pm$ 150 &  0.041 $\pm$ 0.020 &  3.55 $\pm$ 0.27   \\
USco 109           &8.05   &   -2.18 $\pm$ 0.14 &   2851 $\pm$ 150 &  0.054 $\pm$ 0.020 &  4.03 $\pm$ 0.25   \\
USco 128           &8.60   &   -2.40 $\pm$ 0.14 &   2729 $\pm$ 150 &  0.039 $\pm$ 0.015 &  4.02 $\pm$ 0.27   \\
CFHT-Pl 17              &10.36   &   -3.09 $\pm$ 0.17 &   2598 $\pm$ 150 &  0.047 $\pm$ 0.025 &  4.72 $\pm$ 0.25   \\
\\
\textit{Field dwarfs:}  & & & & &  \\
GJ 402             &8.56   &   -2.35 $\pm$ 0.12 &   3125 $\pm$ 120 &  0.121 $\pm$ 0.050 &  4.75 $\pm$ 0.22   \\
Gl 569A            &7.04  &   -1.62 $\pm$ 0.05 &   3548 $\pm$ 120 &  0.350 $\pm$ 0.030 &  4.69 $\pm$ 0.09   \\
Gl 569Ba           &11.18 &   -3.41 $\pm$ 0.05 &   2449 $\pm$ 120 &  0.059 $\pm$ 0.015 &  5.07 $\pm$ 0.13   \\
Gl 569Bb           &11.69 &   -3.62 $\pm$ 0.05 &   2355 $\pm$ 120 &  0.075 $\pm$ 0.010 &  5.32 $\pm$ 0.12   \\
BRI 1222-1221      &11.39 &   -3.50 $\pm$ 0.07 &   2355 $\pm$ 120 &  0.057 $\pm$ 0.015 &  5.08 $\pm$ 0.14   \\
LHS 2243          &10.94\tablenotemark{b} &   -3.31 $\pm$ 0.14 &   2608 $\pm$ 120 &  0.073 $\pm$ 0.010 &  5.21 $\pm$ 0.17   \\
LHS 2397a          &11.15 &   -3.40 $\pm$ 0.05 &   2477 $\pm$ 120 &  0.061 $\pm$ 0.010 &  5.09 $\pm$ 0.11   \\
LHS 2924           &11.63 &   -3.60 $\pm$ 0.05 &   2344 $\pm$ 120 &  0.071 $\pm$ 0.015 &  5.27 $\pm$ 0.11   \\
2MASSI J0825196+211552     &14.97  &   -4.65 $\pm$ 0.10 &   1372 $\pm$ 120 &  0.065 $\pm$ 0.020 &  5.36 $\pm$ 0.23   \\
2MASSI J1029216+162652   &12.64\tablenotemark{b}   &   -3.89 $\pm$ 0.16 &   2100 $\pm$ 120 &  0.068 $\pm$ 0.015 &  5.48 $\pm$ 0.21   \\
2MASSW J1049414+253852    &10.43\tablenotemark{b}  &   -3.12 $\pm$ 0.14 &   2801 $\pm$ 120 &  0.075 $\pm$ 0.015 &  5.12 $\pm$ 0.17   \\
2MASSW J1239272+551537    &13.50\tablenotemark{b}  &   -4.19 $\pm$ 0.16 &   1800 $\pm$ 120 &  0.075 $\pm$ 0.025 &  5.42 $\pm$ 0.24   \\
2MASSW J1444171+300214    &11.11\tablenotemark{b}  &   -3.38 $\pm$ 0.14 &   2543 $\pm$ 120 &  0.076 $\pm$ 0.010 &  5.25 $\pm$ 0.17   \\
2MASSW J1707183+643933   &11.45\tablenotemark{b}   &   -3.52 $\pm$ 0.14 &   2355 $\pm$ 120 &  0.062 $\pm$ 0.015 &  5.16 $\pm$ 0.19   \\
2MASSI J2234138+235956    &11.62\tablenotemark{b}  &   -3.58 $\pm$ 0.14 &   2350 $\pm$ 120 &  0.073 $\pm$ 0.015 &  5.30 $\pm$ 0.19   \\
SDSSp 162414.37+002915.6  &15.34  &   -5.21\tablenotemark{c}  $\pm$ 0.16 &    950 $\pm$ 120 &  0.048 $\pm$ 0.030 &  5.17 $\pm$ 0.38   \\
\\
\textit{Averages} & & & & & \\
for cluster & 7.51 $\pm$ 1.31 & -1.95 $\pm$ 0.52 & 2743 $\pm$ 163 & 0.052 $\pm$ 0.029 & 3.68 $\pm$ 0.42  \\ 
for field   &11.54 $\pm$ 2.04 & -3.52 $\pm$ 0.82 & 2345 $\pm$ 613 & 0.088 $\pm$ 0.071 & 5.17 $\pm$ 0.21  \\  
\\
\textit{Giants:}  & & & & & \\
BD +14 2020 & - & - & 3635 $\pm$ 320 & 1--5 & 0.5 $\pm$ 0.5 \\ 
HD 113285  & -5.34 & 3.30 $\pm$ 0.42 & 2900 $\pm$ 320 & 1--5 & 0.0 $\pm$ 0.5 \\ 
HD 126327 &  -6.56 & 3.78 $\pm$ 0.40 & 3200 $\pm$ 320\tablenotemark{d} & 1--5 & 0.0 $\pm$ 0.5 \\
\enddata

\tablenotetext{a}{Mean values from 3 temperature scales and 3 sets of tracks (\S~\ref{evtrm}). Error estimates are described in \S~\ref{err}}
\tablenotetext{b}{Derived from M$_{J}$-SpT relationship for field dwarfs from \citet{Dahn02}}
\tablenotetext{c}{Luminosity adopted equal to that of Gl229B}
\tablenotetext{d}{\citet{Perrin98} measured 2786 $\pm$ 46 K for this star} 
\end{deluxetable}   

\clearpage

\begin{deluxetable}{lrrrrrr}
\tabletypesize{\small}
\tablecaption{\label{ind} Definition of indices considered in this work ($\mu$$m$)}
\tablewidth{0pt}
\tablehead{
\colhead{name} & \colhead{PC1} & \colhead{PC2} & \colhead{F1} & \colhead{F2} & \colhead{PC3} & \colhead{PC4}}
\startdata
\\
FeH & 1.1890&    1.1960 &   1.1925 &   1.2300 &   1.2265 &  1.2335 \\
\ion{K}{1} & 1.2340 & 1.2410 & 1.2375 & 1.2575  & 1.2540 & 1.2610  \\
H$_{2}$O    &    1.3340 &  1.3380   &  &   &    1.3200 &   1.3240             \\
\ion{Na}{1}     &     2.1950 &   2.2030 &   2.1990 &   2.2150 &   2.2110 &  2.2190 \\
\ion{Ca}{1}   &       2.2525 &   2.2575 &   2.2625 &    2.2675 &   2.2725 &  2.2775 \\
CO       &  2.2960 &  2.2980    &      &       &    2.2860 &   2.2880     \\ 
\\  
\enddata

\tablecomments{H$_{2}$O and CO indices have been defined as the ratio of fluxes averaged
between pseudo-continuum regions PC1-PC2 and PC3-PC4. Larger value of flux ratio corresponds
to weaker absorption feature. Rest indices measure equivalent width of absorption lines.
PC1-PC2 and PC3-PC4 define continuum level while F1 and F2 define the wavelength interval for EW integration. }
\end{deluxetable}

\clearpage

\begin{deluxetable}{lrrrrrr}
\tabletypesize{\small}
\rotate
\tablecaption{\label{indstr}Measurements of indices defined in Table \ref{ind}}
\tablewidth{0pt}
\tablehead{
\colhead{object name} & \colhead{EW(\ion{K}{1}), \AA } & \colhead{EW(FeH), \AA } &
\colhead{H$_{2}$O ind} & \colhead{EW(\ion{Na}{1}), \AA} & \colhead{EW(\ion{Ca}{1}), \AA} & \colhead{CO ind}}
\startdata
\textit{Cluster members:} & & & & & & \\
IC348-355              &     3.48 $\pm$ 1.58 &  - &    0.766 $\pm$ 0.025 &  - &  - & - \\
IC348-363            &     1.82 $\pm$ 1.13 &  - &    0.781 $\pm$ 0.018 &  - &  - & - \\
 CFHT-BD-Tau2           &  - &  - & - &  - &  - & - \\
 CFHT-BD-Tau4          &     3.19 $\pm$ 0.18 & 0 $\pm$ 0.83 & 0.789 $\pm$ 0.003 & 1.63 $\pm$0.48 & 0.23 $\pm$ 0.25 & 0.910 $\pm$ 0.013 \\
 SOri 27          &  - &  - &    0.747 $\pm$ 0.020 &  - &  - & - \\
CRBR 31            &  - &  - &    0.820 $\pm$ 0.059 &  - &  - & - \\
  GY 5       &     2.54 $\pm$ 0.87 & 2.47 $\pm$ 4.13 & 0.869 $\pm$ 0.014 & 1.98 $\pm$ 0.37 & 0.67 $\pm$ 0.19 & 0.911 $\pm$ 0.010 \\
 GY 141       & 6.99 $\pm$ 1.52 &     4.15 $\pm$ 7.18 & 0.746 $\pm$ 0.024 &  - &  - & - \\                
WL 14      &  - &  - & - &     2.73 $\pm$ 0.33 &     1.32 $\pm$ 0.17 &    0.921 $\pm$ 0.009 \\
 2MASSW J1139511-315921         &     5.19 $\pm$ 0.49 & 6.66 $\pm$ 2.33 & 0.728  $\pm$ 0.008 & 1.25 $\pm$ 0.38 & 0.18 $\pm$ 0.20 & 0.874 $\pm$ 0.011 \\
USco 66       &     3.62 $\pm$ 0.30 & 0 $\pm$ 1.41 & 0.864 $\pm$ 0.005 & 3.07 $\pm$ 0.55 & 0.59 $\pm$ 0.28 & 0.900 $\pm$ 0.015 \\
USco 67           &     3.44 $\pm$ 0.49 & 0.82 $\pm$ 2.33 & 0.832 $\pm$ 0.008 & 1.65 $\pm$ 0.65 & 1.08 $\pm$ 0.34 & 0.896 $\pm$ 0.018 \\
 USco 75            &     3.48 $\pm$ 0.59 & 4.29 $\pm$ 2.79 & 0.830 $\pm$ 0.009 &  - &  - & - \\
 USco 85     &  - &  - & - &     2.99 $\pm$ 0.91 &     0.00 $\pm$ 0.47 &    0.854 $\pm$ 0.025 \\
USco 100     &     3.14 $\pm$ 0.28 & 1.76 $\pm$ 1.30 & 0.820 $\pm$ 0.004 & 2.91 $\pm$ 0.52 & 0.72 $\pm$ 0.27 & 0.902 $\pm$ 0.015 \\
USco 109    &     5.02 $\pm$ 0.38 & 3.90 $\pm$ 1.78 & 0.834 $\pm$ 0.006 &  - &  - & - \\
USco 128   &     5.85 $\pm$ 0.46 &  3.23 $\pm$ 2.17 &  0.777 $\pm$ 0.007 &  - &  - & - \\
CFHT-Pl 17  &  - &  - &    0.812 $\pm$ 0.030 &  - &  - & - \\
\\
\textit{Field dwarfs:}  & & & & & & \\
 GJ 402         &     2.10 $\pm$ 0.23 & 0 $\pm$ 1.09 & 0.974 $\pm$ 0.004 & 5.71 $\pm$ 0.38 & 2.16 $\pm$ 0.19 & 0.921 $\pm$ 0.010 \\
Gl 569A       &     0.78 $\pm$ 0.34 & 0.94 $\pm$ 1.63 & 0.969 $\pm$ 0.005 & 5.41 $\pm$ 0.49 & 3.00 $\pm$ 0.25 & 0.927 $\pm$ 0.014 \\
Gl 569Bab      &    12.41 $\pm$ 0.35  & 18.15 $\pm$ 1.66 & 0.750 $\pm$ 0.005 & 2.90 $\pm$ 0.48 & 0.30 $\pm$ 0.25 & 0.839 $\pm$ 0.013 \\
  BRI 1222-1221        &    13.42 $\pm$ 2.28 &  - &    0.783 $\pm$ 0.036 & 4.71 $\pm$ 0.61 & 0.64 $\pm$ 0.31 & 0.825 $\pm$ 0.017 \\
 LHS 2243         &     7.77 $\pm$ 0.20 & 11.30 $\pm$ 0.94 & 0.826 $\pm$ 0.003 & 4.54 $\pm$ 0.45 & 0.02 0.23 & 0.875 $\pm$ 0.013 \\
LHS 2397a       &    12.22 $\pm$ 0.43 & 15.96 $\pm$ 2.01 & 0.798 $\pm$ 0.007 & 2.94 $\pm$ 0.38 & 0.05  $\pm$ 0.20 & 0.859 $\pm$ 0.011 \\
LHS 2924         &    13.55 $\pm$ 0.31 & 19.20 $\pm$ 1.45 & 0.745 $\pm$ 0.005 &  4.46 $\pm$ 1.19 & 0.76 $\pm$ 0.61 & - \\
 2MASSI J0825196+211552          &     4.73 $\pm$ 0.84 &  - &    0.641 $\pm$ 0.013 &     0.32 $\pm$ 0.75 &     0.20 $\pm$ 0.39 &    0.800 $\pm$ 0.021 \\
 2MASSI J1029216+162652         &    18.28 $\pm$ 1.07 & 29.35 $\pm$ 5.03 & 0.686 $\pm$ 0.017 & 1.58 $\pm$ 0.48 & 0.45 $\pm$ 0.25 & 0.787 $\pm$ 0.013\\
2MASSW J1049414+253852         &     5.73 $\pm$ 0.40 & 6.87 $\pm$ 1.91 & 0.856 $\pm$ 0.006 & 6.22 $\pm$ 0.27 & 0.74 $\pm$ 0.14 & 0.900 $\pm$ 0.008 \\
 2MASSW J1239272+551537          &    13.65 $\pm$ 0.98 &    20.13 $\pm$ 4.62 &    0.569 $\pm$ 0.015 & 0 $\pm$ 0.50 & 0.74 $\pm$ 0.25 & 0.800 $\pm$ 0.014 \\
2MASSW J1444171+300214     &     8.97 $\pm$ 0.55 & 12.03 $\pm$ 2.59 & 0.777 $\pm$ 0.009 & 3.33 $\pm$ 0.37 & 0.14 $\pm$ 0.19 & 0.862 $\pm$ 0.010 \\
2MASSW J1707183+643933        &    13.61 $\pm$ 0.67 & 15.80  $\pm$ 3.15 & 0.763 $\pm$ 0.010 & 3.99 $\pm$ 0.66 & 0.76 $\pm$ 0.34 & 0.844 $\pm$ 0.018 \\
2MASSI J2234138+235956       &  - &  - & - &     4.69 $\pm$ 0.31 &     0.12 $\pm$ 0.16 &    0.876 $\pm$ 0.009 \\
SDSSp 162414.37+002915.6          &     7.68 $\pm$ 3.33 &     6.98 $\pm$ 15.7 &    0.118 $\pm$ 0.052 &  - &  - & - \\
\\
\textit{Giants:} & & & & & & \\
 BD +14 2020          &     0.49 $\pm$ 0.37 & 0.32 $\pm$ 1.75 & 0.982 $\pm$ 0.006 & 3.60 $\pm$ 0.52 & 1.88 $\pm$ 0.27 & 0.785 $\pm$ 0.015 \\
  HD 113285    &     4.33 $\pm$ 0.78 & 0 $\pm$ 3.69 & 0.866 $\pm$ 0.012 & 5.04 $\pm$ 1.06 & 2.54  $\pm$ 0.54 & 0.769 $\pm$ 0.029 \\
 HD 126327        &  - &  - & - &     3.62 $\pm$ 0.64 &     1.84 $\pm$ 0.33 &    0.789 $\pm$ 0.018 \\
\enddata

\tablecomments{
``0'' indicates where measured EW was negative but small; 
``--'' indicates that either feature was completely buried in the noise
or that we did not have J or K spectrum. \\
Errors described in \S \ref{errewfr}} 
\end{deluxetable}

\clearpage

\end{document}